\pdfoutput=1
\documentclass[reprint,
 amsmath,amssymb,
 aps,
 onecolumn,
notitlepage,10pt]{revtex4-1}
\usepackage{color}
\usepackage{chemarr}
\usepackage{epsf,mathtools}
\usepackage{graphicx}
\usepackage{dcolumn}
\usepackage{bm}
\usepackage{subfig}
\usepackage{caption}
\usepackage[font=small,labelfont=bf]{caption}
\newcommand{\lt}{<}
\newcommand{\gt}{>}
\usepackage[finalnew]{trackchanges}
\addeditor{}

\begin{document}
\title{\change{Modeling the emergence of ordered polarity patterns in meristemic auxin transport}
{Modeling the emergence of polarity patterns for the intercellular transport of auxin in plants}}
\author{Silvia Grigolon $^{1,\ast}$, Peter Sollich $^{2,\dagger}$, Olivier C. Martin $^{3, \ddagger}$}
\affiliation{$^1$ LPTMS - Laboratoire de Physique Th\'{e}orique et Mod\`{e}les Statistiques, Univ Paris - Sud, UMR CNRS 8626, 15, Rue Georges Cl\'{e}menceau, 91405 Orsay CEDEX, France,\\
$^2$ Department of Mathematics, King's College London, Strand, London, WC2R 2LS UK,\\
$^3$ G\'en\'etique Quantitative et Evolution - Le Moulon, 
INRA/Univ Paris-Sud/CNRS/AgroParisTech, Ferme du Moulon, F-91190 Gif-sur-Yvette, France}
\email{silvia.grigolon@lptms.u-psud.fr}

\email{$^{\dagger}$ peter.sollich@kcl.ac.uk}

\email{$^{\ddagger}$  olivier.martin@moulon.inra.fr}

\begin{abstract}
\remove{Morphogenesis in plants is initiated in meristems, the cells there differentiating under the 
influence of signals from}
The hormone auxin\remove{Auxin} is actively transported throughout 
\change{the meristem}{plants} via protein machineries including the 
dedicated transporter known as PIN. 
\remove{In the last decade it has been discovered that}
The associated transport is \change{highly polarised}{ordered} with nearby cells driving auxin flux in 
similar directions. Here we provide a model of both the auxin transport and of
the dynamics of cellular polarisation based on flux sensing. Our main findings are:
(i) spontaneous \add{intracellular} PIN polarisation arises 
\change{if and only if cooperative effects in}{if}
PIN recycling \change{are strong enough}{dynamics are sufficiently non-linear}, (ii) there is no
need for an auxin concentration gradient, and (iii)
\change{orderly}{ordered} multi-cellular patterns of PIN polarisation are favored by molecular noise.
\end{abstract}

\maketitle

\section{Introduction}

In plants, the initiation of different organs such as roots, leaves or flowers
\remove{arises in regions called meristems}
\remove{These regions contain stem cells
that have the potential to proliferate and differentiate into any
of the plant's organs throughout the whole life of the plant.
The choice of which organ is initiated}
depends on the cues received
by \remove{the} cells, be\remove{-} they from the environment or \remove{as} signals produced by the plant itself \cite{scheres2001}.
Amongst these signals, the hormone auxin plays a central role.
Auxin was discovered over a half century ago
along with some of its macroscopic effects on leaf and root growth \cite{went}. 
\change{Its local accumulation leads to cell differentiation and
commitment to the initiation of new organs. Like a number of other molecular
species, auxin}{It} is actively transported
\change{inside meristems and the resulting distribution of
this hormone is ultimately key to}{throughout the whole plant
and it is a major driver of} the plant's architecture  \cite{reinhardt2000, reinhardt2003}. 

In the past decade, much
has been learned about
the molecular actors controlling auxin movement\remove{in meristems}. 
First, cell-to-cell auxin fluxes depend on two classes of 
transporters \cite{palme1999, muday2001, petrasek2009}:
(i) PIN, \add{(for `PIN-FORMED'')}, that pumps auxin from inside to outside 
cells \cite{galweiler1998}, and
(ii) AUX1, \add{(for `AUXIN RESISTANT 1'')}, \change{that}{which} pumps auxin from outside to inside cells. 
Second, auxin accumulation drives cell proliferation and differentiation.
\remove{observations have shown that auxin accumulates according to near
deterministic patterns.}
Third, cells are polarized \change{for}{in terms of} their PIN content, that is PIN transporters
localize mainly to one side of cells \cite{traas2013}. In addition, these polarizations are 
similar from cell to cell so that auxin is systematically transported along the
direction of this polarization. That \change{\emph{ordering}}{ordering} has major consequences for the
\add{growth and} morphogenesis \change{induced within the meristem}{of the plant} because
it affects the distribution of auxin, and auxin drives 
\add{both organ growth and} the initiation of
\remove{promordia which themselves produce the} new organs \cite{auxinbook, pinbook}.
Much work has focused on how PIN polarization patterns lead to auxin distributions, but
two major questions remain unanswered concerning 
the emergence of PIN polarization patterns:
(i) \add{How} can PIN become polarized in cells in the absence of auxin gradients? 
(ii) Can PIN polarization patterns be coherent on the scale of many cells?

To address these questions, we take a modeling approach here, incorporating 
the main ingredients of what
is currently known about (i) intercellular auxin transport and (ii)
intracellular PIN dynamics. 
We will first provide a deterministic framework using differential
equations for modeling the dynamics of auxin and of PIN cellular polarization.
Our model exhibits multiple steady states \change{which}{that} we characterize,
the simplest ones being translation-invariant with all cells having the same 
\change{polarization for PIN}{PIN polarization}. 
We find that the emergence of polarization depends on 
the degree
\change{of cooperative effects}{of non-linearity} 
within the PIN recycling dynamics.
We then include molecular noise in this system coming from the 
stochastic dynamics of PIN intra-cellular localization. Interestingly,
for biologically realistic values of the \remove{model's} parameters
\add{of the model}, the 
system is driven into a state where cells coherently polarize in the same
direction. In effect, the noise selects \remove{for} a robust self-organised state 
having homogeneous PIN polarization, corresponding to a noise-induced ordering
scenario. 

\section{The model of auxin dynamics and PIN recycling}

\begin{figure*}
\centering
\includegraphics[scale=0.5]{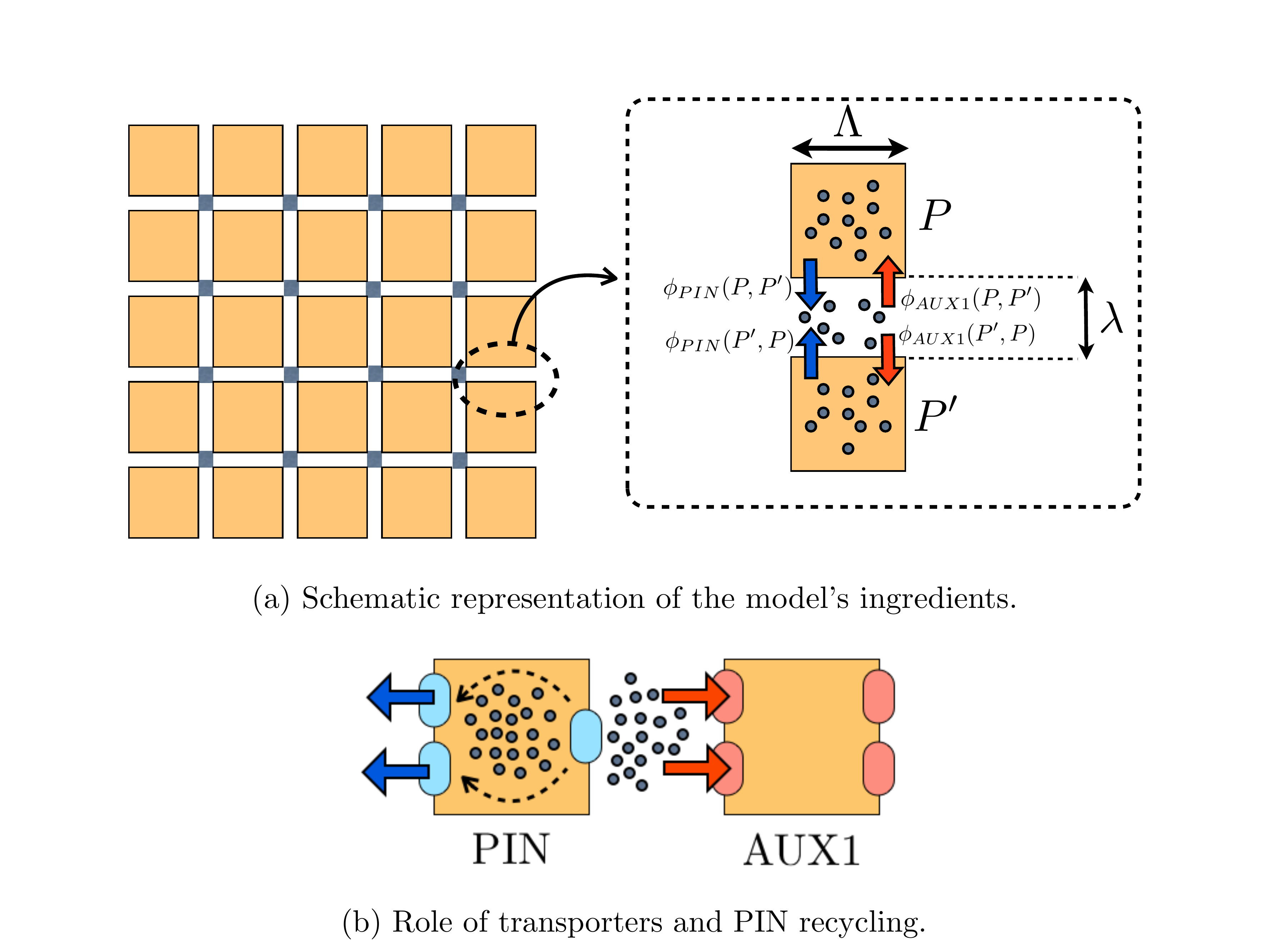}
\caption{(a) Schematic two-dimensional view of the system consisting of a single layer of cells. Cubic cells of size $\Lambda$ are in orange, apoplasts of width $\lambda$ are in white. On the right: a zoom on two neighboring cells. Gray circles stand for auxin and red arrows represent the incoming AUX1 mediated fluxes in cells, while light blue arrows represent the outgoing PIN mediated fluxes. In these views from above, the thickness ($\Lambda$) 
of the cells is not shown. (b) Role of transmembrane transporters. PINs pump auxin from the inside of the cell to an adjacent apoplast. AUX1 plays the reverse role, pumping auxin from apoplasts to the inside of the cell. Dashed arrows within a cell illustrate PIN recycling.}
\label{fig:schematicmodel}
\end{figure*}

Auxin transport \change{in shoot meristems takes place mainly in
a sheet that consists}{in plants is typically organized
in sheets, each sheet consisting} of a single layer of cells. 
\change{and}{For instance in the case of the tips of shoots,
almost all the transport arises in a single cell sheet} which is referred to as the L1 layer \cite{dereuille2006, buylla2010, vernoux2010, smith2006, jonsson2006}. \add{For our model,} we shall therefore work 
\change{using a single}{with one} layer of cells. \change{to
represent the meristem. Our model of 
the Shoot Apical Meristem (SAM)}{We} start with a lattice of cubic cells having edges of
\add{length} $\Lambda$ separated by 
apoplasts -- the space between two adjacent cells -- of width $\lambda$ (\emph{cf.} Fig.~\ref{fig:schematicmodel} which represents
a view from above of this system). Tissues
consist of closely packed cells so $\lambda \ll \Lambda$; typical values are
$\lambda \approx 1 \mu m$ and $\Lambda \approx 20 \mu m$.
\remove{It has long been known that} The hormone auxin \cite{reinhardt2000}
\remove{is an important morphogen
for plant development and more specifically that it is the key driver of meristemic dynamics . It} is 
subject to different processes:
\begin{itemize}
\item production and degradation inside cells, with rate constants $\beta~$ and $\rho$;
\item passive diffusion within cells, within apoplasts, and also between cells and apoplasts;
\item active transport across cell membranes via transporters.
\end{itemize}
Since cell membranes form barriers to exchanges of molecules, taking a molecular species from one side to the other
often requires dedicated transporters. In the case of auxin, the cell membrane does allow some
amount of diffusion of the hormone but much less than the inside of the cell or of the apoplast where diffusion is very rapid. We call $D$ the associated diffusion constant (measured in $\mu \text{m}^2/\text{s}$) \emph{within} the
membrane of thickness $\epsilon$ whereas formally we consider diffusion inside cells and inside apoplasts to arise infinitely quickly; as a consequence, intra-cellular variations of auxin concentration are negligible and so are those within an apoplast. In addition, auxin is subject to active processes that transport it
across the cell membrane. Experimental 
evidence has shown that there are different molecular transporters for the in-going and out-going fluxes, transporters called respectively AUX1 and PIN \cite{reinhardt2003, friml2006}. These transporters are normally localised on the cell membrane where they can play their role to actively transport auxin between the inside and the outside of the cell. 
The out-going transporters belong to a large family whose members specialize to different organs and tissues of the 
plant \cite{krecek2009}: in our context, we will refer to these transporters simply as PIN \cite{paponov2005}.
\remove{PIN1 being the best studied member.} 

The dynamics of auxin concentration in each region (cell or apoplast) is \remove{then} 
specified by the trans-membrane \change{fluxes of auxin}{flux densities of auxin
($\phi_{AUX1}$ and $\phi_{PIN}$ for the active transport
and a diffusion contribution proportional to the diffusion constant $D$)} along with production 
and degradation terms. In the case of cells, we have
\begin{equation}
\begin{split}
\frac{d A_c(P,t)}{dt}&=\beta~ - \rho A_c(P,t) + \Lambda^{-1} 
\sum_{P'} [ \phi_{AUX1}(P,P',t) - \phi_{PIN}(P,P',t) + D (A_a(P,P',t) - A_c(P,t))/\epsilon ].
\end{split}
\label{eq:auxindifferentialequation}
\end{equation}
In this equation, $A_c(P,t)$ is the auxin intra-cellular concentration of the cell centered at
position $P=(x,y)$ at time $t$,
and $A_a(P,P',t)$ is the auxin concentration in the apoplast separating nearest neighbor
cells $P$ and $P'$. Both concentrations will be specified in micro-moles 
per liter ($\mu$M for micro-molar). Furthermore, only
the diffusion constant of auxin \emph{within} the cell membrane appears because it is far smaller than
that within a cell or apoplast; \add{note that the flux is proportional to the gradient, 
thus the factor $D/\epsilon$ where $\epsilon$ is the thickness of the membrane}. 
$\phi_{AUX1}(P,P',t)$ and $\phi_{PIN}(P,P',t)$ are the auxin flux 
\emph{densities} carried by the transporters through the \add{corresponding} ``face''
\add{of cell $P$}, \emph{i.e.}, 
the area of the membrane of 
cell $P$ that faces cell $P'$. By convention,
the sign of each flux is positive, the one for PIN going from the inside to the outside of the cell, and
the one for AUX1 going from outside to inside\remove{ the cell}. These flux densities have units of
micro-moles per second per surface area ($\mu m^2$). The sum over cells
$P'$ is restricted to the neighbors of $P$ so in effect one sums
over all sides of the cell $P$ under consideration that connect \add{it} to the rest of the system. The 
parameters $\beta$ and $\rho$ are the rates of auxin production and degradation. In addition,
the \add{division by the} factor $\Lambda$ \add{(the width of a cell)} appears because one goes 
from flux densities \remove{across a cell face} to effects on the concentrations inside cells. Lastly, in our framework as depicted in Fig.~\ref{fig:schematicmodel}, apoplasts connect only to 
cells and vice versa, \add{so} there are \change{no}{neither} 
cell-to-cell \remove{contacts} nor apoplast-to-apoplast contacts.

In a similar fashion, the concentration $A_a(P,P',t)$ of auxin in the apoplast 
\add{(of thickness $\lambda$)} separating 
cells $P$ and $P'$ obeys the differential equation:
\begin{equation}
\begin{split}
\frac{d A_a(P,P',t)}{dt} &=  \lambda^{-1} [ \phi_{PIN}(P,P',t) - \phi_{AUX1}(P,P',t)+ \phi_{PIN}(P',P,t) - \phi_{AUX1}(P',P,t)+\\
&+ D (A_c(P,t) + A_c(P',t) - 2 A_a(P,P',t) )/\epsilon ] ~ .
\end{split}
\label{eq:auxindifferentialequation_apoplast}
\end{equation}
Note that there is neither production nor degradation of auxin in the apoplast (it is a passive medium
and auxin has a long lifetime in the absence of the active degradation processes present in cells).

In \emph{Arabidopsis}, which currently is the most studied plant, \change{flux due to the}{propensity
of} AUX1 influx transporters seems to be several times higher than that \change{due to}{of} passive diffusion \cite{kramer2006, kramer2008}, 
thus active processes are probably the main drivers of auxin
distribution. Furthermore, the transporters AUX1 and PIN are believed to be completely unidirectional; the associated molecular mechanisms
are unclear but involve first the binding of
auxin and then conformational changes. Because these processes are analogous 
to enzymatic reactions, we model
the associated auxin fluxes via irreversible Michaelis-Menten kinetics:
\begin{equation}
\phi_{AUX1}(P,P',t)= \frac{N^{AUX1}}{\Lambda^2} \cdot \alpha \cdot \frac{A_a(P,P',t)}{1+\frac{A_a(P,P',t)}{A^*}+\frac{A_c(P,t)}{A^{**}}},
\label{eq:fluxaux1}
\end{equation}
\begin{equation}
\phi_{PIN}(P,P',t) = \frac{N^{PIN}}{\Lambda^2} \cdot \gamma \cdot \frac{A_c(P,t)}{1+\frac{A_a(P,P',t)}{A^*}+\frac{A_c(P,t)}{A^{**}}},
\label{eq:fluxPIN}
\end{equation}
where $\alpha$ and $\gamma$ are kinetic constants analogous to catalysis rates. The 
factor $\Lambda^2$ 
on the \change{left}{right}-hand side of these equations corresponds to 
the surface of the face of each cell, and connects the flux \emph{density} to the (absolute) flux. 
At a molecular level, 
$N^{AUX1}$ (respectively $N^{PIN}$) refers to the \emph{number} of AUX1 
(respectively PIN) transporters on the area of $P$'s membrane which
faces cell $P'$. Finally, 
$A^{*}$ and $A^{**}$ play the role of Michaelis-Menten constants associated with saturation effects; 
these could have been taken to be different in Eqs.~\ref{eq:fluxaux1} and \ref{eq:fluxPIN} without any qualitative consequences for the \change{model's behavior}{behavior of the model}. 

We are not aware of any experimental evidence that the distribution of AUX1 transporters changes with time 
or that these transporters contribute to cell polarity. Thus we shall assume that their numbers 
are constant on each face of the cell. In contrast, PIN transporters are particularly important for 
driving morphogenesis through the formation of polarity patterns. Often they 
define \change{clear}{clearly} polarized fields in tissues \cite{dereuille2006, sassi2013} where cells see 
their PINs predominantly localised to one of their faces, \add{with} the \add{specific} face being the same for 
many cells. That polarity leads to coherent auxin transport, 
\change{accumulation of auxin in 
specific regions, and then the first signs of committement to differentiation in the
form of primordia .}{even on the
scale of the whole plant, allowing in particular auxin to be
transported from shoots to roots}\cite{friml2006-2, benkova2003}. To take into
account this possibility of 
\change{forming polarized distributions of}{intracellular polarization of} PIN, we introduce the four 
faces of a cell as 
\change{$R$ for right, $U$ for upper, $L$ for left and $D$ for down}
{$N$ for north, $S$ for south, $E$ for east and $W$ for west}. (The 
two faces parallel to the \change{L1 layer}{sheet} play no role in our simplified 
\change{modeling}{model} \add{involving a single layer of cells}.) Then each face of a cell has 
a potentially variable number of PIN transporters:
\begin{center}
$N_f^{PIN}$, \, \change{$f=R, U, L, D$.}{$f = N, S, E, W$.}
\end{center}
Furthermore we impose the constraint
$\sum_{f} N_f^{PIN} \equiv \sigma$, a cell-independent constant so each cell has the same
number of PIN transporters at all times.

The dynamics of PIN seem complex: it is known that PINs are subject to ``recycling'' within a cell through different mechanisms including transport from the membrane to the \change{golgi}{Golgi} apparatus and back to the membrane \cite{petrasek2009, lofke2013, tanaka2013, kleinevehn2011}. Most modeling takes PIN dynamics to be driven by surrounding auxin concentrations \cite{vanberkel2013, kramer2008, kramer2006, smith2006, jonsson2006, wabnik2010}. For instance it has been postulated that PIN might accumulate to the membrane facing the neighboring cell with the highest concentration of auxin \cite{smith2006, jonsson2006}. As a consequence, the presence of an auxin gradient becomes a necessary condition for PIN polarization. Here we consider instead  dynamics based on \textit{flux sensing} where PIN recycling rates are modulated by the amount of auxin flux transported by those same PIN transporters. Mathematically, we take the PIN dynamics on a face $f$ 
\change{($f=R, U, L, D$)}{($f=N, S, E, W$)} of a cell to be 
\change{spec! ified as follows}{specified by a Hill equation 
of exponent $h$}:
\begin{equation}
\begin{split}
\frac{d N_f^{PIN}}{dt}&=- \frac{3}{4 \tau} N_f^{PIN} \frac{1}{1+\bigg(\frac{\phi_f^{PIN}}{\phi^*}\bigg)^h}+\frac{1}{4 \tau} \sum_{f'} N_{f'}^{PIN} \frac{1}{1+\bigg(\frac{\phi_{f'}^{PIN}}{\phi^*}\bigg)^h} .
\end{split}
\label{eq:pinequation}
\end{equation}
\add{Note that PINs are treated here as continuous variables because in the following the number of molecules is high so such an approximation is appropriate. Nevertheless,
later we shall treat the actual numbers via our stochastic model.} 
\change{These}{The above} differential equations model the competitive recycling of 
the PINs amongst the faces of a 
given cell in a flux dependent manner. Note that $\tau$ is the characteristic time scale of 
the recycling process and that the dynamics enforce the 
constraint of conservation of the total number of PIN transporters inside each cell. Also, in 
Eq.~\ref{eq:pinequation},
$\phi_f^{PIN}$ is the flux \change{$\Lambda^2 \phi_{PIN}$ (not a flux density) flowing}{density} through face $f$
while $\phi^*$ is a \add{Michaelis-Menten-like} constant.

To understand the consequences of Eq.~\ref{eq:pinequation}, consider first 
the low flux limit, $\phi_f^{PIN}$ being small for all faces, so
that all denominators can be ignored. At a molecular level, each transporter leaves the cell membrane at 
a rate proportional to $\tau^{-1}$; transporters are then brought into the cytoplasm or 
\change{golgi}{Golgi} apparatus; finally they get reallocated randomly 
to any of the 4 faces. In such a low flux regime, cells will show no PIN polarization. 
Second, consider instead the case where for at least one face $f$,
$\phi_f^{PIN}/\phi^{*}$ is large. That face will \emph{benefit} from the recycling, recruiting more
transporters from the other faces than it \change{looses}{loses}.
($\phi^{*}$ is just the scale at which flux sensing in this system becomes important). At an
individual transporter level, the competitive recycling means that 
transporters which are actively shuttling auxin see their rate of recycling go down. How this happens 
depends on unknown molecular details, nevertheless, the rate of detachment of a transporter
from the membrane probably depends on what fraction of the time it is binding auxin 
and thus the rate of detachment will show a dependence on $\phi_f^{PIN}$. One 
may attempt to model this via a simple 
hyperbolic law to describe saturation effects. To be \remove{still} 
more general, we
have \remove{further} introduced a Hill exponent, $h$, 
\add{into the dynamics as given in} Eq.~\ref{eq:pinequation}. 
Such a functional
form is often used \change{for modeling binding rates}{in the kinetic 
modeling of binding processes; in that framework,}
$h$ is an integer related
to the number of molecules that must co-localize, 
\add{and as such, it reflects} \remove{but more generally $h$ 
can be used to simply parametrize} co-operative effects. In the absence of detailed knowledge of the molecular mechanisms
controlling PIN recycling,
we use this phenomenological form
\add{where $h$ is associated with the non-linearity
of the PIN recycling dynamics} and 
\add{we} will see whether or not $h$ plays an important role.
\add{In the Supplementary Material, we will see that our conclusions
are insensitive to the precise form of the equations describing PIN recycling by replacing
the Hill form with a stretched exponential.}

The model is now completely specified and involves the 15 parameters 
$\Lambda$, $\lambda$, $\epsilon$, $\alpha$, $\beta~$, $\gamma$, $\rho$, 
$A^*$, $A^{**}$, $\phi^{*}$, $\tau$, $\sigma$, 
$N^{AUX1}$, $D$, and $h$. Some of these parameters can be absorbed in scale changes. 
Nevertheless, to keep the physical  interpretation as
transparent as possible, we stay with the dimensional form of the equations. Whenever possible, we assign
values to the parameters using published estimates or compilations 
thereof~\cite{Deinum_Thesis, kramer2006, kramer2008}. 
For instance, mass-spectrometry measurements \cite{ljung2001} 
in very young leaves quantify the concentration of auxin to be about 250 picograms per milligram of tissue.
Since the molecular mass of auxin is about 175 \add{Da}, $A_c$ is of the order of 1 micro-molar. 
As we shall see, in the steady-state regime, $\beta / \rho = A_c$, a relation providing a 
constraint on those two parameters. Furthermore, a direct estimate of $\beta$
follows from isotopic labeling measurements \cite{ljung2001}
which show that biosynthesis replenishes auxin within about one day; we have thus
set $\beta$=1/day. \add{Radioactive labeling has also provided estimates for mean displacement
velocities of auxin} \cite{kramer2006, kramer2008} \add{which we have used to constrain 
the parameters $\alpha$ and $\gamma$}. Unfortunately, for other parameters 
\add{(and in particular the Michaelis-Menten constants)} no \change{such}{direct
or indirect estimations from} experimental
data \change{is}{are} available. For most such cases, 
we use ball-park estimations that seem reasonable, 
\add{for instance
a hundred PIN molecules seems like a too low number while $10^4$ is perhaps on the high side}.
However, for $h$, which provides a phenomenological parameterization 
of \change{cooperative}{non-linear} effects
in PIN recycling, we have little choice but to study the behavior of the model as a function of
its value. We use the same strategy for $D$. Thus, both 
$D$ and $h$ will be used as control parameters, allowing us to map out a 
two-dimensional phase diagram. For instance, when increasing $D$,
passive diffusion will overcome the effects of active transport, allowing one to probe the 
importance of active \emph{vs}.\ passive transport in the establishment of PIN polarization.  
Unless specified otherwise, all other parameter values are set as provided in Table~\ref{tab:parameters}.
\begin{table}
\begin{tabular}{| c | c |}
\hline
\hline
$\beta~$ & 1 $\mu$M $\cdot$ day$^{-1}$\\
$\rho$ & 1 day$^{-1}$\\
$\alpha$ & 0.1 L $\cdot$ s$^{-1}$\\
$\gamma$ & $10^{-4}$ L $\cdot$ s$^{-1}$\\
$A^*$ & 2 $\times10^{-3}$ $\mu$M\\
$A^{**}$ & 0.8 $\mu$M\\
$\phi^*$ & 4 $\times 10^{-6}$ moles $\cdot$ $\mu$m$^{-2}$ s$^{-1}$\\
$N^{AUX1}$ & 200 per face\\
$\sigma$ & 1000\\
$\Lambda$ & 20 $\mu$m\\
$\lambda$ & 1 $\mu$m\\
$\epsilon$ & 10 nm\\
$\tau, \tau_{1D}$ & 30 min\\ 
\hline
\hline
\end{tabular}
\caption{Parameters used in the model. M stands for \textit{molarity}, \emph{i.e.}, number of moles per liter.
L stands for liter.}
\label{tab:parameters}
\end{table}

Since our aim is to understand how \change{organised}{ordered} 
polarity patterns arise in a system described by this model,
it is appropriate to define \remove{something like} an order parameter to quantify the ordering of flux directions 
or PIN intra-cellular localisation. We thus introduce the two-dimensional 
\textit{polarization} vector $\vec{\delta}$ for a cell at position $P=(x,y)$; its components depend on the face-to-face difference of the number of PINs along each direction (say horizontal or vertical) in the following way:
\begin{equation}
\vec{\delta}(x,y) \equiv \begin{cases}
\delta_1(P)=\frac{N_E^{PIN}(P) - N_W^{PIN}(P)}{\sigma}\\
\\
\delta_2(P)=\frac{N_N^{PIN}(P) - N_S^{PIN}(P)}{\sigma}
\end{cases}
\label{eq:polarization}
\end{equation}
The vector in Eq.~\ref{eq:polarization} has a \change{magnitude}{length} $|\vec{\delta}(P)| \in [0,1]$: the two \change{extremal}{extreme} values represent respectively the unpolarized case, \emph{i.e.}, 
$N_f^{PIN} = \sigma/4$ for all $f$, and the fully polarized case, \emph{i.e.}, $N_f^{PIN} = \sigma$ for one face while $N_f^{PIN} = 0$ for all other faces. Individual components can vary in $[-1,1]$ and the \change{extremal}{extreme} points give the maximum polarization in one direction or the other.

A first step will consist in understanding the behavior of this system in a one-dimensional framework. 

\section{Analysis of the one-dimensional model}
\subsection{Dynamical equations}

Let us replace \change{the L1 layer,}{the square lattice} represented in Fig.~\ref{fig:schematicmodel} \remove{via a square lattice,}
by a \emph{row} of cubic cells forming a one-dimensional lattice. As before, between two
\change{adjacents}{adjacent} cells there is exactly one apoplast. 
\remove{Starting with Eq.~1, we see that the up and down terms no longer arise if we force all diffusion and transport to arise between cells and apoplasts. Defining $\Delta$ as the distance 
between two nearest-neighbor cells ($\Delta = \Lambda+\lambda$),one has:}
\remove{
where the positions $P$ and $P'$ of Eq.~1 have been replaced by 
$x$ and $x' = x \pm \Delta$. Similarly,
for apoplasts one has}
In this one-dimensional model, \add{all diffusion and
transport is horizontal and} PIN is defined only on the left (West) and right (East) face of each cell. 
\add{The dynamics of $A_c$ and $A_a$ in each cell are obtained
from Eqs. 1 and 2 by setting the vertical fluxes to 0.}
Given the constraint of conservation of total PIN transporters in each cell \change{(}{and the fact that} only two faces contribute\remove{)}, 
the dynamics of PIN numbers are completely determined via the dynamics of $N^{PIN}_E$ and there is just one independent equation for each cell:
\begin{equation}
\frac{d N^{PIN}_E}{dt}=-\frac{1}{\tau_{1D}} N^{PIN}_E \frac{1}{1+(\frac{\phi^{PIN}_E}{\phi^*})^h}+\frac{1}{\tau_{1D}}N^{PIN}_W \frac{1}{1+(\frac{\phi^{PIN}_W}{{\phi^*}})^h}
\label{eq:pindifferential1D}
\end{equation}
\add{where $\tau_{1D} = 4 \tau / 3$ and}
\remove{In this equation,} $N^{PIN}_W = \sigma - N^{PIN}_E$
\remove{and a factor 2 arising from using 
Eq.~1 has been absorbed into $\tau$}.
Furthermore, polarization is no longer a vector but a scalar, given by the first 
component of Eq.~\ref{eq:polarization}. It varies in $[-1,1]$: when $\delta(x) \approx -1$, almost all the PINs are on the left-hand side of the cell, while when $\delta(x) \approx 1$ they are almost all on the right-hand side.

\subsection{Steady-state auxin concentrations given translation-invariant PIN configurations}

Assuming periodic boundary conditions, the row of cells becomes a ring; this idealization is convenient for the mathematical and phase diagram analysis. Consider the steady-state solutions of the differential equations. With periodic boundary conditions, one expects some steady states to be translationally invariant. In that situation, all quantities are identical from cell to cell and from apoplast to apoplast. We can then drop all time and spatial dependence in the variables,
e.g., $A_c(P,t) = A_c$ for all $P$ and $t$. 

\begin{figure}
\includegraphics[scale=0.5]{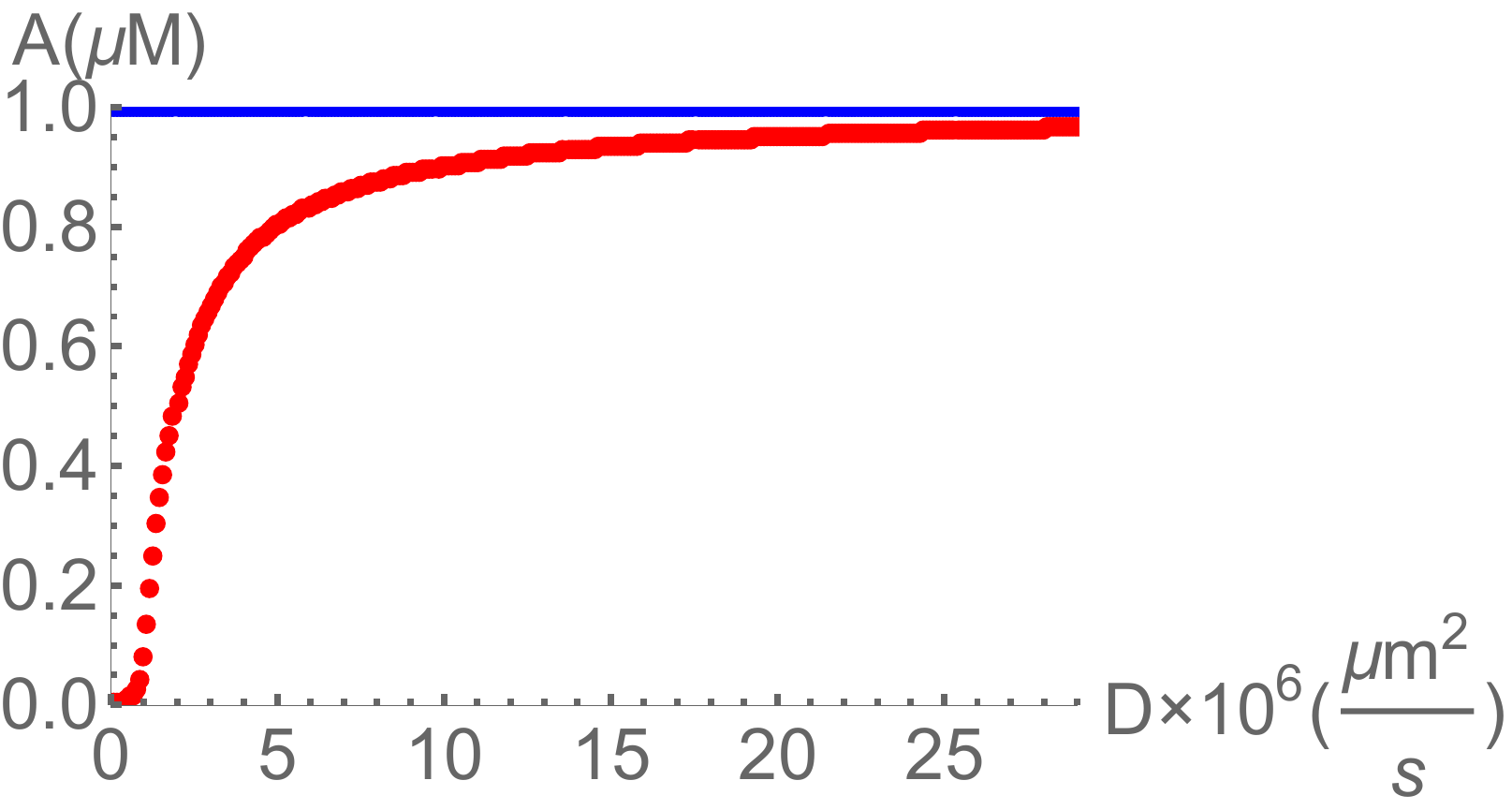}
\caption{Auxin steady-state concentrations (red for apoplasts and blue for cells) in an arbitrary
PIN translation-invariant configuration as a function of the diffusion constant $D$ in $\mu \text{m}^2/\text{s}$. 
The other relevant model parameters are given in Table~\ref{tab:parameters}.}
\label{fig:auxinvsDA}
\end{figure}

We consider here an arbitrary translation-invariant configuration of PIN transporters (steady-state
or not), \add{which implies that the auxin equations will depend only on the total number of transporters per cell.} One \add{then} has the following equations 
for steady-state auxin concentrations:
\begin{equation}
\left\{
\begin{array}{l}
0=\beta~-\rho A_c+\frac{2D}{\Lambda \epsilon} (A_a-A_c)+\frac{2 \alpha}{\Lambda^3} N^{AUX1} \frac{A_a}{1+\frac{A_a}{A^*}+\frac{A_c}{A^{**}}}- \frac{\gamma}{\Lambda^3} \sigma \frac{A_c}{1+\frac{A_a}{A^*}+\frac{A_c}{A^{**}}}\\
\\
0=\frac{2 D}{\lambda \epsilon} (A_c - A_a)- \frac{2 \alpha}{\lambda \Lambda^2} N^{AUX1} \frac{A_a}{1+\frac{A_a}{A^*}+\frac{A_c}{A^{**}}}+\frac{\gamma}{\lambda \Lambda^2} \sigma \frac{A_c}{1+\frac{A_a}{A^*}+\frac{A_c}{A^{**}}}\\
\end{array}
\right..
\end{equation}
\remove{Note that because we imposed translation invariance, the equations do not depend on the particular values of $N_f^{PIN}$ because all that matters here is the total number of PIN transporters per cell. } These two equations
determine $A_c$ and $A_a$. \change{Since}{One can first
solve for $A_c$ by noting that} in apoplasts there is no source or degradation of auxin,
\change{in}{thus} in the steady state the total flux (transport and diffusion counted 
algebraically) through an apoplast vanishes and \add{so} the same holds
for cells. \change{Thus}{Therefore} within cells
auxin degradation must compensate exactly auxin production, leading to $\beta-\rho A_c=0$. 
This result, namely $A_c = \beta/\rho$, is independent of all other parameters and in particular of
the PIN polarization and of $D$ as illustrated in Fig.~\ref{fig:auxinvsDA}. Furthermore, 
\change{since $A_c$ is now known, the second equation}{with $A_c$ determined in this way the two equations become equivalent and} can be solved for $A_a$.

Experimental evidence~\cite{kramer2006, kramer2008} suggests that active transport dominates passive (diffusive) transport in \textit{Arabidopsis}. Thus the biologically relevant regime probably
corresponds to $D$ small. In
the low diffusion limit ($D \to 0$), the last equation shows that $A_a$ goes to a limiting
value that is strictly positive.(A zoom of Fig.~\ref{fig:auxinvsDA} would also show this is
the case.) Solving this equation using the values of the parameters 
in Table~\ref{tab:parameters}, 
one finds that the concentration of auxin in cells is much greater than that in apoplasts because
$2 N^{AUX1} \alpha \gg \sigma \gamma$, \emph{i.e.}, auxin molecules are more easily transported 
by AUX1 than by PIN (\emph{cf}. the left limit in Fig.~\ref{fig:auxinvsDA}). One may also consider what happens when 
diffusion is important; clearly as $D \to \infty$, the transporters become irrelevant and 
the equations immediately show that $A_c$ and $A_a$ become equal. The overall behavior
is displayed in Fig.~\ref{fig:auxinvsDA}.

\subsection{Translation-invariant dynamics of PIN in the quasi-equilibrium limit for auxin}

Microscopic molecular events associated with auxin transport (be \remove{-}they active or passive) 
arise on very short time scales whereas the PIN recycling requires major cellular
machinery and so arises on much longer time scales. Let us therefore take the \emph{quasi-equilibrium} 
limit where auxin concentrations \remove{are fast variables and thus can be assumed to} 
take on their steady-state values
\remove{given the PIN configuration. 
Furthermore, we take the numbers of PIN transporters to be translation invariant,
so we can use the results for auxin concentrations} $A_c$ and $A_a$\remove{ of the previous section}. 
\remove{These being PIN independent, the auxin concentrations in fact just stay at their exact steady-state values.} 
Consider now the 
dynamical equation for $\delta$, the PIN polarization. 
\remove{This equation can be derived easily 
from the equation for $N^{PIN}_E$ and using the constraint on PIN number conservation.}
Since it
involves a single variable, it can always be written as gradient descent relaxational dynamics,
\emph{i.e.}, there exists a function $\mathcal{F}(\delta)$ such that:
\begin{equation}
\frac{d \delta}{dt}=- \frac{1}{\tau_{1D}} \frac{d \mathcal{F}(\delta)}{d \delta} .
\end{equation}
$\mathcal{F}(\delta)$ plays the role of an effective potential. $\mathcal{F}(\delta)$ is minus the 
\change{antiderivative}{integral} of a known function of $\delta$; this \change{antiderivative}{integral} can be obtained in closed form\change{, namely}{ in terms of hypergeometric functions (see Supplementary Material).}
\remove{
where $_2F_1[a,b,d,z]=\sum_{k=0}^{\infty} \frac{(a)_k (b)_k}{(d)_k k!} z^k$, $(a)_k$ is the Pochhammer symbol and 
$c=\frac{\gamma \sigma}{\phi^*}\frac{A_c}{1+\frac{A_a}{A^*}+\frac{A_c}{A^{**}}}$.
One can check that $c$ scales with the inverse of the diffusion constant. 
}

\begin{figure*}
\includegraphics[scale=0.5]{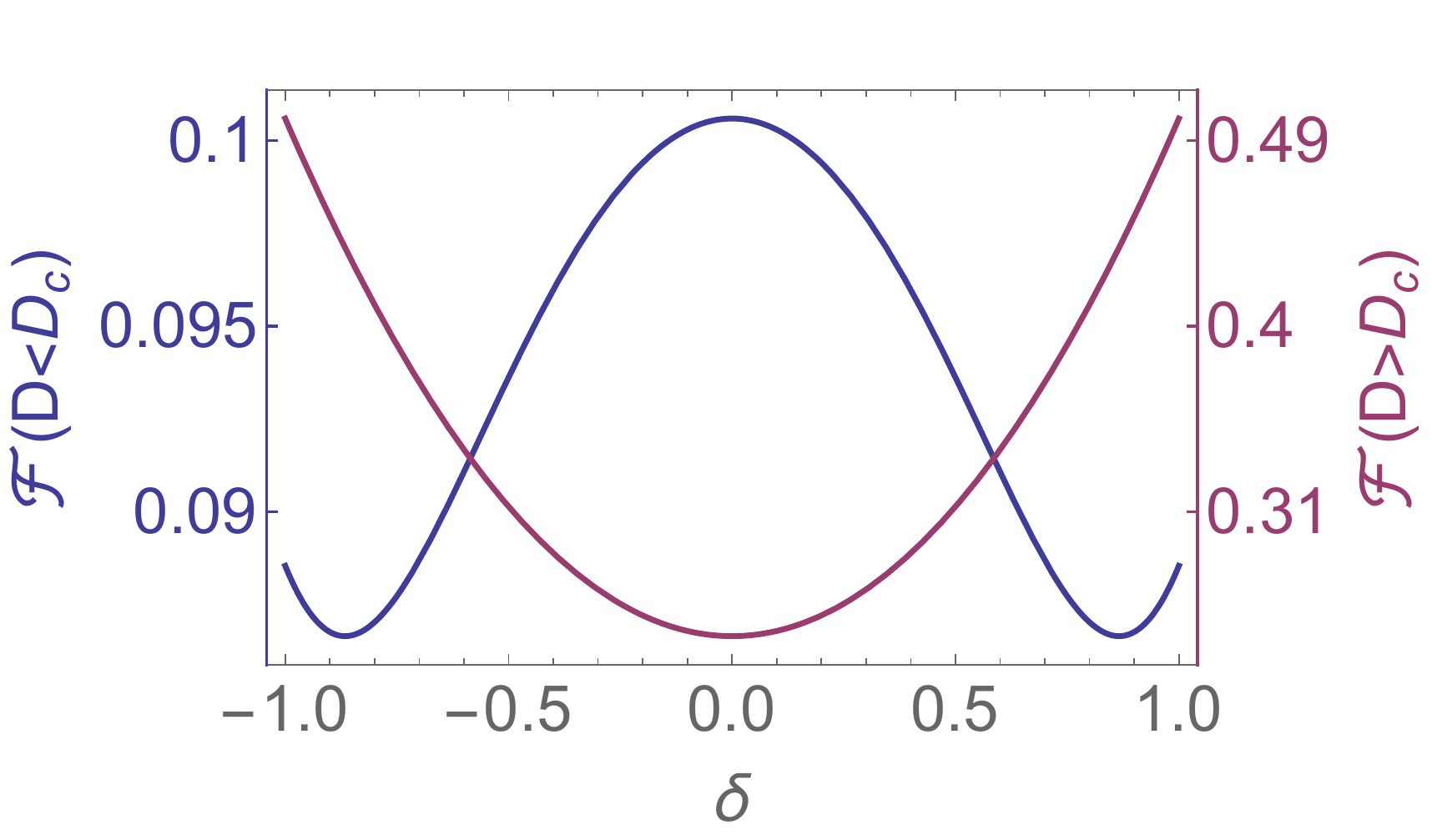}
\caption{Effective potential as a function of the polarization $\delta$ for $D>D_c$ (purple) 
and $D<D_c$ (blue) rescaled with $\sigma^2$.
$h=2$, other parameter values are given in Table~\ref{tab:parameters}.}
\label{fig:doublewellh2}
\end{figure*}

The extrema of $\mathcal{F}$ correspond to steady states for the PIN dynamics, 
\emph{i.e.}, $d \delta / dt = 0$. Maxima are unstable and minima are stable.
Thus it is of interest to map out the form of $\mathcal{F}$ as a function
of the parameter values. Take for instance $h=2$. Starting with a large value for $D$,
use of Mathematica shows that
$\mathcal{F}$ has a single global minimum, corresponding to 
the unpolarized steady state, $\delta = 0$. Then as $D$ is lowered,
\add{(the Mathematica code of the Supplementary Material provides the user with a knob to change $D$),} $\mathcal{F}$
takes on a double-well shape, symmetric
about the zero polarization abscissa where the curvature is now negative. 
At the same time, two new local minima appear at $\pm \delta^*$. 
Thus as $D$ is lowered, the unpolarized state becomes unstable while two new stable 
steady states of polarization $\pm \delta^*$ appear; this situation is illustrated
in Fig.~\ref{fig:doublewellh2}.
If we had used instead $h=0.5$, we would have found no regime in $D$ where $\mathcal{F}$ 
has the double-well structure: there, the only steady state is the unpolarized one.

Within this framework, it is possible to determine a critical point $D_c$, {\emph i.e.}, 
a threshold value of the diffusion constant, below which spontaneous
symmetry breaking sets in. The value of $D_c$ is obtained from the following condition
(see also Supplementary Material):
\begin{equation}
\frac{\partial^2 \mathcal{F}}{\partial \delta^2} |_{\delta=0}=0.
\label{eq:criticalequation}
\end{equation}
In particular, for $h=2$, this leads to a critical 
value $D_c \approx 9.4 \times 10^{-7} \mu \text{m}^2/\text{s}$.
This overall framework provides a convenient intuitive picture for PIN dynamics.
\remove{Unfortunately
no such effective potential can be obtained in the two-dimensional model as will be discussed later.}

\subsection{Spontaneous symmetry breaking and phase diagram for translation-invariant steady states}

\begin{figure}[b!]
\includegraphics[scale=0.8]{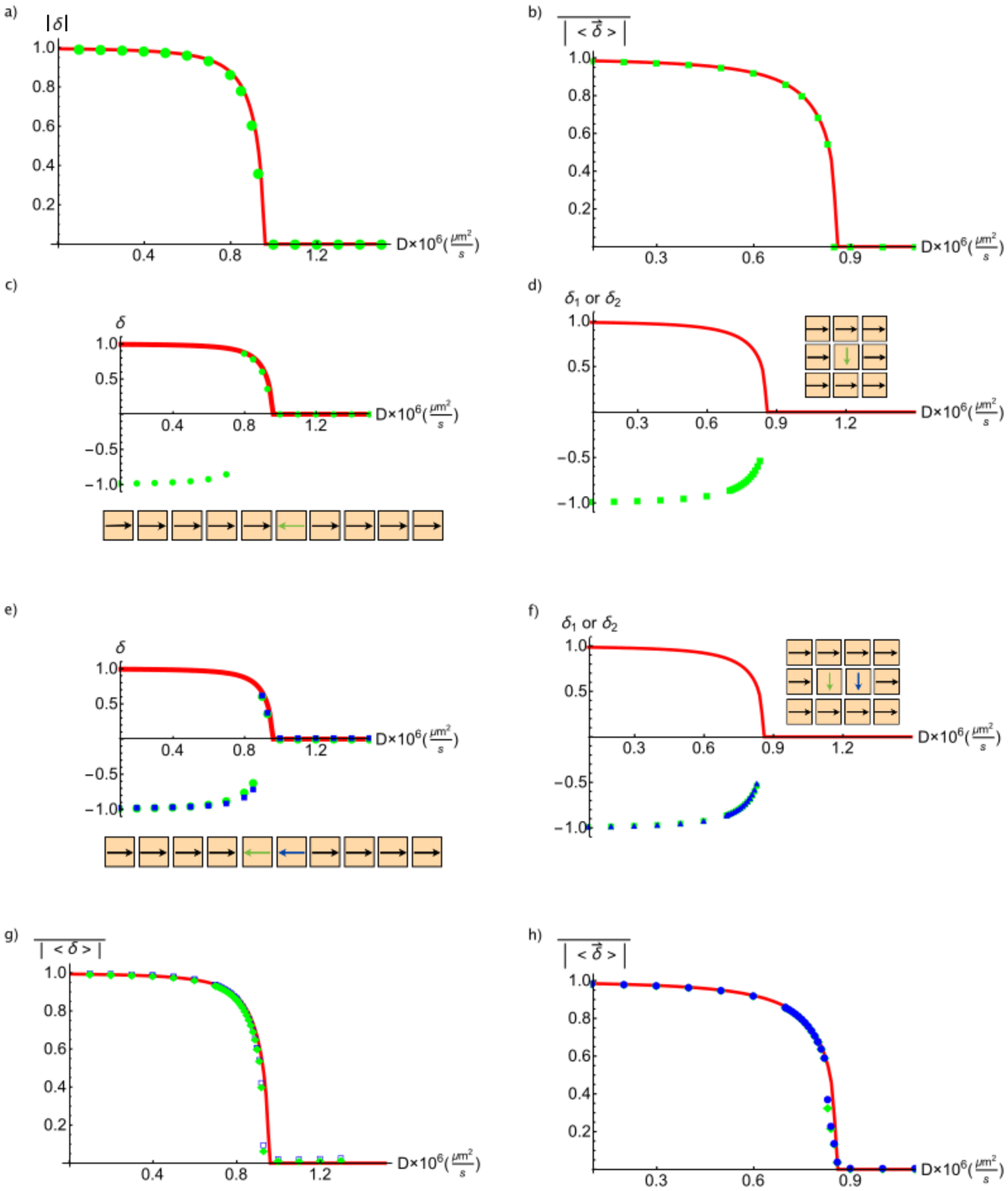}
\caption{}
\label{fig:phasevsDA}
\end{figure}

The previous formalism is complicated because of the form of $\mathcal{F}$. However if one is
only interested in the steady states and one does not care about relaxational dynamics, the 
steady-state equation
to solve is relatively simple (\emph{cf.} Eq.~\ref{eq:pindifferential1D} where the
left-hand side must be set to 0)\remove{ and there is no need 
to appeal to a
quasi-equilibrium approximation}.
We assume as before that $N_E^{PIN}$ is translation invariant but also that it is subject to 
PIN recycling. Since $A_c$ and $A_a$ in steady states
have been previously calculated, all quantities
in Eq.~\ref{eq:pindifferential1D} are known except for $N_E^{PIN}$; it is enough then to
solve the associated non-linear 
equation (we have used Mathematica for \change{that}{this purpose}). At given $h$ not too small, we
find a transition from polarized to unpolarized states as $D$ crosses the threshold $D_c$ (see
Fig.~\ref{fig:phasevsDA}a which illustrates the case $h=2$).
\addtocounter{figure}{-1}
\begin{figure} [t!]
  \caption{a) and b) Absolute value of the (translation-invariant) PIN polarization as a function of the diffusion 
constant $D$ in $\mu \text{m}^2/\text{s}$ in steady states respectively for the 1D and the 2D models.
Red line: analytical result \add{obtained} using Mathematica. Green circles: results of simulating the dynamics of the model
containing respectively 20 cells on a ring and 20$\times$20 cells on a lattice until a steady state was reached; a fourth-order 
Runge-Kutta algorithm \cite{numericalrecipes} was used and
starting configurations were randomized but had positive local PIN polarizations. c),e),d),f) PIN polarization at a defect (green and blue) and in the absence of a defect (red) for a ring of 20 cells
as a function of $D$ for the 1D case (c) and e)) and a lattice of 20$\times$20 cells for the 2D case (d) and f)). Drawings (below and insets): initial orientation of PIN polarizations; the green and blue arrows represent
the defects.  g) and h) Absolute value of the mean PIN polarization per site, averaged over time, as a function of \add{the} diffusion \add{constant} for \remove{both} the stochastic model for three different ring/lattice sizes (in g), $N_{cells}=20$ green diamonds, $N_{cells}=10$ blue squares, while in h) $N_{cells}=5$ blue circles, $N_{cells}=10$ green diamonds) and for the deterministic model (red line). Simulations were performed 
using cells on a ring/lattice.  $D_c$ is slightly lower when using stochastic dynamics. In all the plots, $h=2$ while other parameter values are given in Table~\ref{tab:parameters}. In g), $\tau=1$s.} 
\end{figure}
The position of the threshold depends on $h$. However, if $h$ is too small, the transition point 
disappears, and
there are no longer any polarized steady states. To illustrate the situation, consider fixing $D$ to a small
value, say $D = 10^{-7} \mu \text{m}^2/\text{s}$, and then solve for $N_E^{PIN}$
as a function of the Hill exponent $h$. For $h$ less than a 
critical threshold $h_c \approx 1.09$, there
is a unique solution and it corresponds to the unpolarized state, $N_E^{PIN} = N_W^{PIN} = \sigma/2$.
For $h > h_c$, two new steady states appear which are polarized. These
two states are related by the left-right symmetry, so \change{one has}{there is} a 
spontaneous symmetry breaking transition
at $h_c$ (see Fig.~\ref{fig:bifurcation1D}). As $h \to \infty$, these states tend towards full
polarization, $\delta = \pm 1$.
\begin{figure}
\includegraphics[scale=0.5]{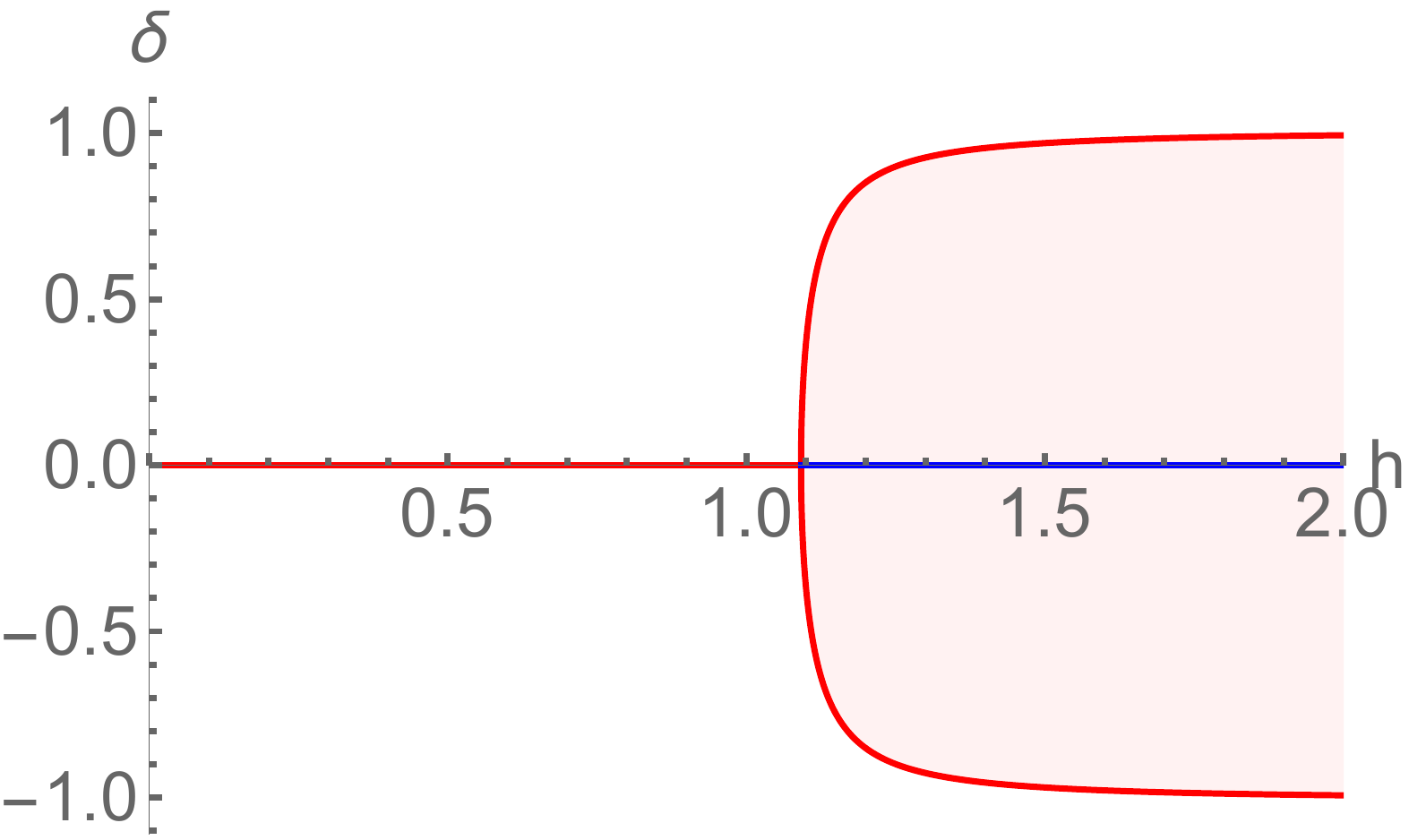}
\caption{Bifurcation diagram for translation-invariant states in the one-dimensional 
model. $\delta$ is the PIN polarization. The unpolarized state is stable for $h \lt h_c \approx 1.09$ (orange). Beyond that threshold, 
two symmetric polarized states appear. These are stable (in red) whereas the unpolarized state becomes unstable (in blue). 
Here $D=10^{-7} \mu \text{m}^2/\text{s}$ while other parameter values are given in Table~\ref{tab:parameters}.}
\label{fig:bifurcation1D}
\end{figure}
To represent simultaneously the behavior as a function of the diffusion constant $D$ and
of the Hill exponent $h$, Fig.~\ref{fig:doublephasevsDA}a provides the overall phase diagram via a heat 
map. Note that when $h$ is too low \emph{or} $D$ is too high, the only steady state is 
the unpolarized one.
\begin{figure}
\includegraphics[scale=0.5]{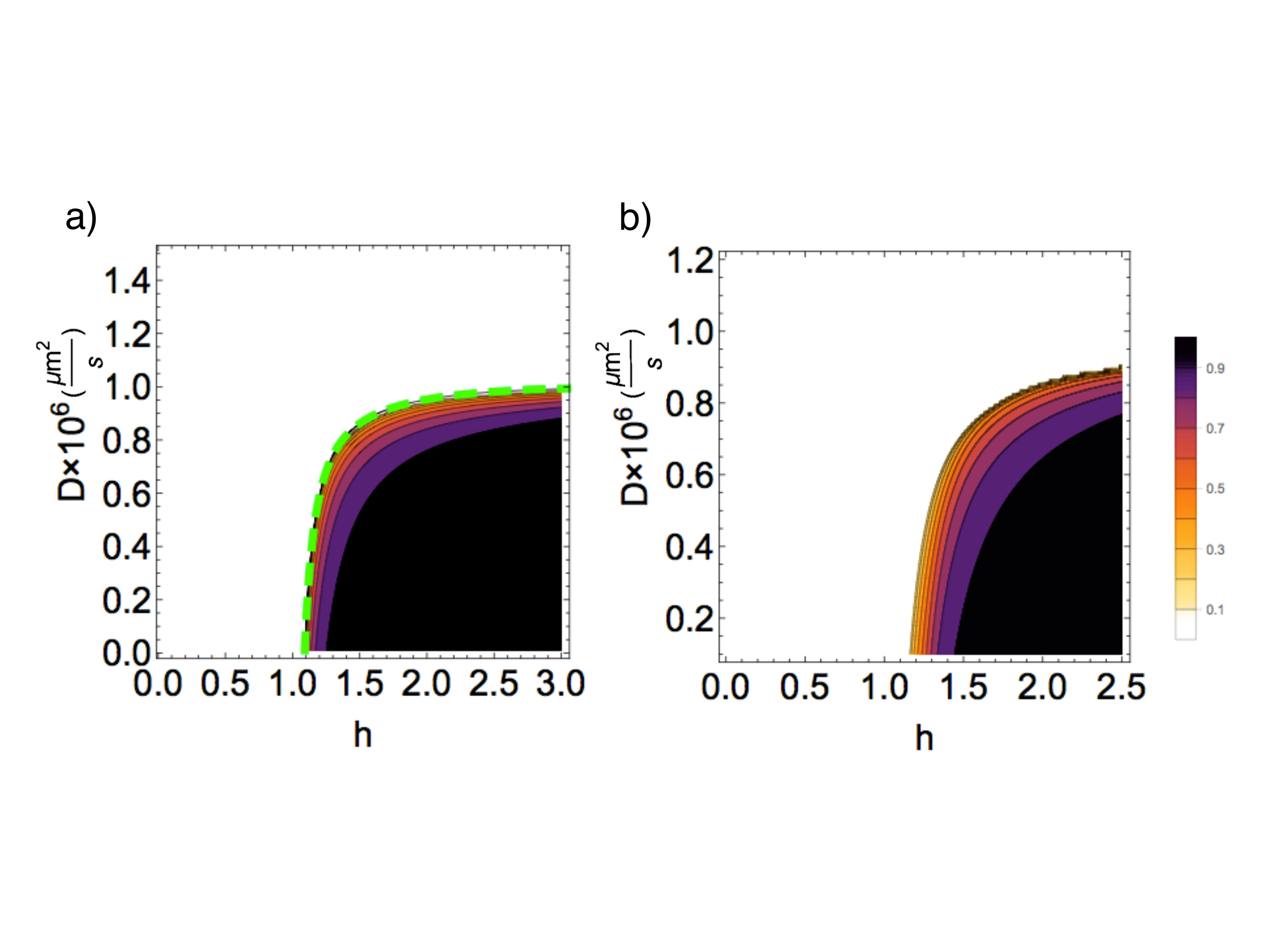}
\caption{Heat map of PIN polarization in translation-invariant steady states as a function of $h$ \add{(dimensionless)} and 
$D$ in $\mu \text{m}^2/\text{s}$ \remove{re}scaled by a factor $10^{6}$ for better readability. Other 
parameter values are given in Table~\ref{tab:parameters}. a) 1D model. The green dashed line refers to the theoretically
derived critical line. b) 2D model.}
\label{fig:doublephasevsDA}
\end{figure}

The origin of this spontaneous symmetry breaking is the change of stability of the
unpolarized state. To quantitatively understand that phenomenon,
set $N^{PIN}_E = \sigma/2 + \delta/2$ and then
linearize Eq.~\ref{eq:pindifferential1D} in $\delta$. Defining
$F = \gamma A_c \sigma / [2 (1 + A_a/A^* + A_c/A^{**}){\Lambda^2} {\phi^*} ]$ (this 
is \remove{a constant} independent of polarization but \remove{which} varies with $D$ because of its
dependence on $A_a$), one has 
$\phi^{PIN}_E /\phi^* = ((1 + \delta)/\sigma) F$ and 
$\phi^{PIN}_W /\phi^* = ((1 - \delta)/\sigma) F$. Then the linearization in $\delta$ leads to: 
\begin{equation}
\tau \frac{d \delta}{dt}= -2 \Big[ \frac{1 - (h-1)F^h}{(1+F^h)^2} \Big] \delta.
\label{eq:role_of_h}
\end{equation}
Instability arises if and only if $(h-1)F^h > 1$. Note that 
the case of Michaelis-Menten type dynamics ($h=1$)
\add{therefore} does \emph{not} lead to PIN polarization. To have spontaneous polarization 
inside cells, \change{one must have \emph{cooperative} 
dynamics with}{the non-linearity must be strong enough. The 
mathematical condition is}
$h > h_c = 1 + F^{-h_c}$ where $h_c$ is the critical Hill exponent where
the instability sets in. This demonstrates the essential role of the non-linearity parametrised here by $h$. 
Of course other forms of non-linearity can be expected to lead to similar conclusions. In particular,  we have found
that the same qualitative behaviour arises when using stretched exponentials rather than Hill-functions (see Supplementary Material). 
We thus conclude that in general the spontaneous polarisation of PIN is driven by the strength of the non-linearity parameterizing the PIN recycling dynamics. 

One may also investigate the stability of the polarized steady state. 
First, within the space of translation invariant configurations, a linear stability analysis
using Mathematica shows that the polarized state is always linearly stable.
This is exactly what the ad\add{i}abatic approximation predicts (\emph{cf.}
Fig.~\ref{fig:doublewellh2}).
Second, one can ask whether our translation-invariant steady states are \emph{global} 
attractors when they are linearly stable. We have addressed this heuristically 
by simulating the dynamical equations starting
from random initial conditions. When $h \le h_c$ (or $D \ge D_c$ if one considers $h$ as fixed), 
it seems that the unpolarized
state is the \emph{only} steady state and that all initial conditions converge to it. When $h \gt h_c$, 
the system \add{always} seems to \remove{always} go to a steady state: we have never observed any 
oscillatory or chaotic behavior.
Sometimes \change{these}{the} steady states are the previously found translation-invariant 
polarized states but sometimes they are not\change{; this 
latter situation
leads us to consider now non translation-invariant steady states}
{, and contain cells with opposite signs for the PIN polarization}.
\add{This situation is much like what happens when quenching
the Ising model where there is a proliferation of such 
disordered states. In the Supplementary Material, we characterize some
of these non translation-invariant steady states. }
The main conclusion to draw from \change{these results}{the arguments gathered there} 
is that as one approaches $D_c$ the number
of steady states diminishes. Furthermore, one expects that this effect is accompanied
by a reduction in both stability and size of basin
of attraction of steady states having defects, leading to an
increase of the coherence length (or domain sizes where a domain is a block of
cells having the same sign of polarization) as one
approaches $D_c$. Such properties naturally lead one to ask whether noise might
enhance the coherence of polarization patterns, driving the emergence
of order from disorder~\cite{helbing2002}.

\subsection{Properties of the stochastic model}

Since the number of \add{PIN} transporter molecules \remove{(PIN and AUX1)} in our 
system is modest, noise in the \add{associated} dynamics
may be important. \add{Thus in this section we reconsider the system
by using a stochastic framework where each individual PIN transporter 
can move from one face to another according to probabilistic laws. 
The parameters of those laws are known via the fluxes in the deterministic model: these
fluxes give the \emph{mean} number of such PIN recycling events per unit time.  
To study the stochastic model, we simulate these random events from which we can
extract the average properties arising in the presence of such molecular noise.
(See the Supplementary Material for implementation details.)}

\remove{To include molecular noise in 
reaction-diffusion systems, it is common practice to use a Lattice Boltzmann model framework
and thermodynamical considerations to render the reaction dynamics
stochastic. However, in the present case, the Michaelis-Menten form of the rates of change of 
molecular concentrations as well as the PIN recycling dynamics
do not correspond to mass action reactions so a different framework is necessary. We thus take a Gillespie-like approach where each process type (production or degradation of auxin inside a cell, transfer
of auxin from one side to the other of a
membrane, or PIN recycling from one cell face to another) is rendered stochastic.
For instance, diffusion of auxin across a face of a cell is associated with two underlying
fluxes, one in each direction. If dt is a short time interval, the noiseless dynamics
specifies that an average number of molecules M = r dt will contribute to the 
underlying flux where r is the rate of that process. But because of the molecular nature of the process,
the true number of molecules generating that flux will be a 
Poisson variable of mean r dt. Applying this rule to all the 
underlying fluxes contributing to the auxin differential equations and to 
Eq. , the stochastic dynamics are naturally 
implemented without the introduction of any new parameters. (Note for instance
that temperature dependencies feed-in only via the deterministic parameters such as D.)
However, there are millions of auxin molecules in a cell and so the associated noise is
negligible. In contrast, the number of PIN transporters in a cell is modest and so
the noise in PIN recycling can a priori be important. Our simulation thus implements
the noise in PIN recycling but not in the other processes.} 

The stochastic dynamics \remove{defined by the procedure just described}
are ergodic, so given enough time the system will thermalize,
there being a unique ``thermodynamic equilibrium state''. Although in principle
this state depends on the value of $\tau_{1D}$, if auxin concentrations are close to 
their steady-state values which is the case here, $\tau_{1D}$ just introduces a
time scale and has no effect on the equilibrium state. We use simulations to study
the equilibrium, \change{putting}{with} a particular focus on the behavior of PIN polarization. Observables must be 
averaged over time. Just as in other thermodynamical systems having spontaneous symmetry
breaking, care \add{then} has to be used when extracting the order parameter. We thus measure
the mean PIN polarization defined by first averaging $\delta$ over all cells to obtain
$\langle \delta \rangle$, then taking the absolute value,
$| \langle \delta \rangle |$, and then averaging over simulation time:
$\overline{| \langle \delta \rangle |}$. 
\remove{In practice we begin with a 
``cold start'', \emph{i.e.}, the starting configuration of the system has all cells similarly 
polarized; then we run to thermalize the system before performing any measurements. }

In Fig.~\ref{fig:phasevsDA}g
we show the \remove{previously defined} mean polarization \add{thus defined} as a function of $D$ for systems having 10 and 20 cells.
At low $D$ the analysis of the model in the absence of noise suggested that the system 
\change{could}{will}
not polarize \add{coherently} because the typical noiseless steady state had random 
polarizations \add{(cf. previous section)}. Nevertheless,
here we see that, in the presence of noise, the system seems to have a global
polarization, in agreement with the order from disorder
scenario~\cite{helbing2002}. If one refers to the special 
translation-invariant steady state in the absence of noise, 
it seems at low $D$ that the presence of
noise leads to almost exactly the same value of the order parameter, so noise can be thought of as
``selecting'' that particular ordered state. As $D$ grows, polarization 
intensity decreases and noise effects are
amplified. As might have been expected, polarization is lost earlier in the presence of noise
than in its absence. 
\remove{To check whether these simulations indeed are in equilibrium, 
we compared the polarization values to those obtained when starting the simulations with
``hot starts'', \emph{i.e.}, the starting configuration of the system having cells randomly polarized.
For D  greater than 5  10$^{-7}$ m$^2$/s the two approaches agreed very well while for lower values of 
D the runs with random initial conditions failed to thermalize well. As a result, only
for intermediate and large values of D can one be confident in the simulation.}

\remove{To follow-up on this last caveat, }Fig.~\ref{fig:phasevsDA}g could
be interpreted as suggesting that the 
equilibrium state in the
stochastic model has a real transition between a polarized phase and an unpolarized
one. However, \remove{thermalization is difficult at low D;} \change{furthermore,}{one has to bear in mind that} for a system containing a large enough 
numbers of cells the equilibrium state will in fact contain multiple domains of polarization,
some being oriented in one direction and others in the opposite direction. This is inevitable
in any one-dimensional system having short range interactions~\cite{landaustatphys, huangstatmech}, and so
no true long-range order arises in this system if the number of cells is allowed to 
be arbitrarily large. To add credence to this claim, note that the polarization curves
are slightly different for the different lattice sizes, the polarization
\emph{decreasing} as the number of cells increases. It is thus plausible that in the 
limit of an infinite number of cells, the polarization vanishes for all $D$.

\section{Analysis of the two-dimensional model}
\subsection{Steady-state auxin concentrations given translation-invariant PIN configurations}

\remove{We now tackle the two-dimensional model (see Fig.~1). 
For each cell there are
four faces contributing to active and passive transport of auxin and 
each allows for PIN recycling as given by Eq.~5. As 
for the one-dimensional model, we begin with the deterministic dynamics,
again taking periodic boundary conditions for the ease of analysis. 
Unless otherwise specified, parameter values are set to those in Table I.}
\change{Let us first focus on}{In two dimensions we again begin by considering} auxin steady-state concentrations in the presence of translation-invariant PIN configurations. 
\remove{This allows us to drop the spatial and temporal dependence of the different variables. 
Thus for all $(x,y)$ labeling
cell positions of the system, one has}
%
%
\add{Auxin concentrations are then also translation-invariant, but compared
to the one-dimensional case, vertical and horizontal apoplasts need not have
the same concentrations of auxin. We denote these concentrations
as $A^{N}_a$ and $A^{W}_a$}.
\remove{where $A^{R}_a(x,y)$ (respectively $A^{U}_a(x,y)$) is the right (respectively up) apoplast 
auxin concentration when considering the cell at position $(x,y)$.}

In all steady states, the total rate of auxin production must be compensated
by the total rate of auxin degradation. 
\remove{For translation-invariant steady states}
This \remove{gives} immediately \add{gives} $A_c = \beta / \rho$ just like in the one-dimensional model. In addition,
$A^{W}_a$ is determined by the equation
\begin{equation}
0=2 D (A_c - A^{W}_a)- 2 \alpha N^{AUX1} \frac{A^{W}_a}{1+\frac{A^{W}_a}{A^*}+\frac{A_c}{A^{**}}} + 
\gamma \sigma^{E}  \frac{A_c}{1+\frac{A^{W}_a}{A^*}+\frac{A_c}{A^{**}}},
\end{equation}
where $\sigma^W=N^{PIN}_E + N^{PIN}_W$. $A^{N}$ is determined by the analogous equation
in which the index $W$ is replaced by $N$ and 
$\sigma^N=N^{PIN}_N + N^{PIN}_S$. Thus, in contrast to the one-dimensional case, the concentration of auxin 
in apoplasts
depends not only on model parameters like $D$ but also on PIN polarization. 
Unpolarized configurations lead to $\sigma^W=\sigma^N=\sigma/2$ and
then $A^{N}_a = A^{W}_a$, in which case the equations take the same form as in one dimension.

\begin{figure*}
\includegraphics[scale=0.5]{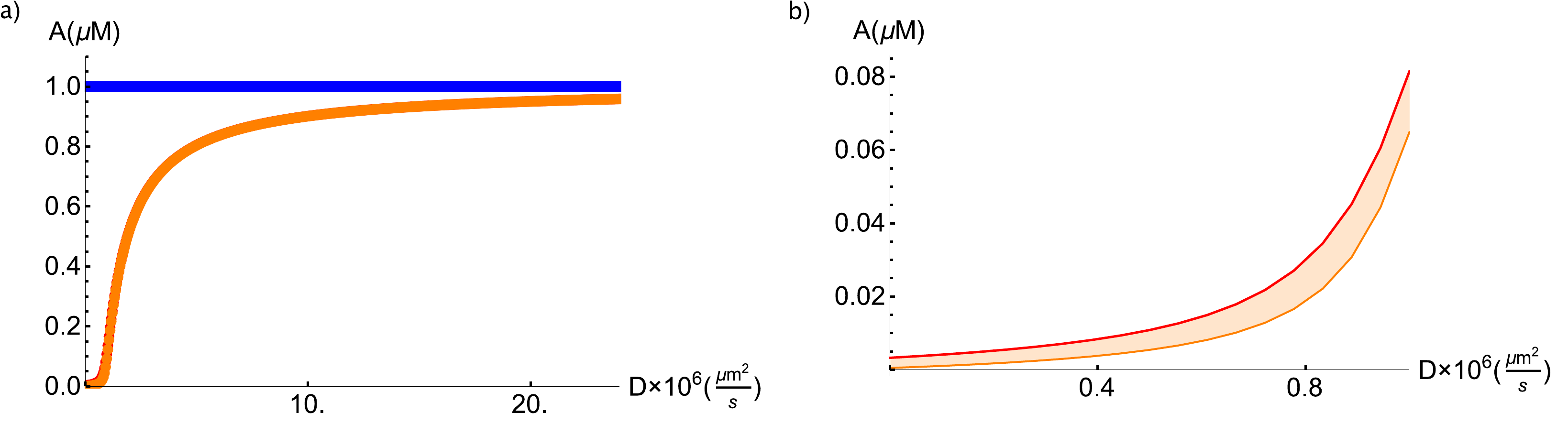}
\caption{Steady-state auxin concentrations in translation-invariant PIN configurations 
as a function of the diffusion constant $D$. On the left: in blue, the concentration
within cells. On the left and via a zoom on the right: in red and orange, the minimum and 
maximum values of auxin concentration within apoplasts,
arising when the number of PIN transporters \change{facing}{on the faces of} the apoplast takes on its minimum and maximum value.
$h=2$, other parameter values as in Table~\ref{tab:parameters}.}
\label{fig:auxinprofile2D}
\end{figure*}

The lowest \add{and highest} possible value of $A^{W}_a$ arises when $\sigma^W=0$ \add{and $\sigma^W=\sigma$, respectively.} \remove{, and the highest value arises when
$\sigma^W=\sigma$} These lower and upper bounds are represented in 
Fig.~\ref{fig:auxinprofile2D} along with the value of $A_c$ as a function of $D$.
Clearly, auxin concentrations are hardly affected at all by PIN polarization. Furthermore, 
both qualitatively and quantitatively, the situation is very close to that in
the one-dimensional model.

\subsection{Translation-invariant dynamics of PIN in the quasi-equilibrium limit for auxin}

In the one-dimensional model, we saw that translation-invariant dynamics
of PIN polarization followed from \change{an effective potential}{a potential energy
function} when auxin
was assumed to be in the quasi-equilibrium state. In two dimensions, there are 
four dynamical variables which satisfy the conservation law
$N^{PIN}_E + N^{PIN}_W + N^{PIN}_N + N^{PIN}_S = \sigma$. Each $N^{PIN}_f$ obeys
a first order differential equation; the question now is whether these follow from
\change{an effective potential}{a potential energy function} $\mathcal{F}$:
\begin{equation}
\frac{\tau d N^{PIN}_f}{dt}=-\frac{\partial \mathcal{F}}{\partial N^{PIN}_f} .
\end{equation}
The answer is negative: no \remove{effective} potential exists because
the velocity field has a non-zero curl. Nevertheless, if in the initial conditions
the PINs obey the symmetry $N^{PIN}_N = N^{PIN}_S$ (or the symmetry $N^{PIN}_E = N^{PIN}_W$), 
then this symmetry is preserved by the dynamics. (Note that the
symmetry is associated with reflecting the system of cells \change{along}{about} an axis.)
Then one sees that the differential equations for the two other PIN numbers are nearly identical to those
in the one-dimensional model. For instance, if $N^{PIN}_N = N^{PIN}_S$, the equation
for $N^{PIN}_E$ is that of the one-dimensional model if one substitutes $\sigma$
by $\sigma^W$. The difficulty is that $\sigma^W$ itself follows from solving
the differential equations and thus can depend on time. Although one does not
have a true \remove{effective} potential \add{energy function}, the important property is that
the instantaneous rate of change of $N^{PIN}_W$ can be
mapped to its value in the one-dimensional model via the aforementioned 
substitution. We thus expect to have the same
kind of spontaneous symmetry breaking where the unpolarized steady state goes from being stable
at low $h$ to being unstable at high $h$, with an associated appearance of stable polarized
steady states.

\subsection{Spontaneous symmetry breaking and phase diagram for translation-invariant steady states}

To determine the translation-invariant steady states, one must solve six simultaneous non-linear equations,
two of which give $A_a^W$ and $A_a^N$ in terms of the $N^{PIN}_f$, the other four being
associated with PIN recycling. We \change{solve these equations} {tackle this task} using Mathematica.

Qualitatively, one obtains the same behavior as in the one-dimensional model. As 
displayed in Fig.~\ref{fig:phasevsDA}b, there is a
continuous transition between a polarized state at low $D$ and an unpolarized state
at large $D$. 
\remove{The qualitative justification is as before, 
\emph{i.e.}, at large $D$ active transport is irrelevant and auxin diffusion leads
to unpolarized cells.}

\begin{figure*}
\includegraphics[scale=0.4]{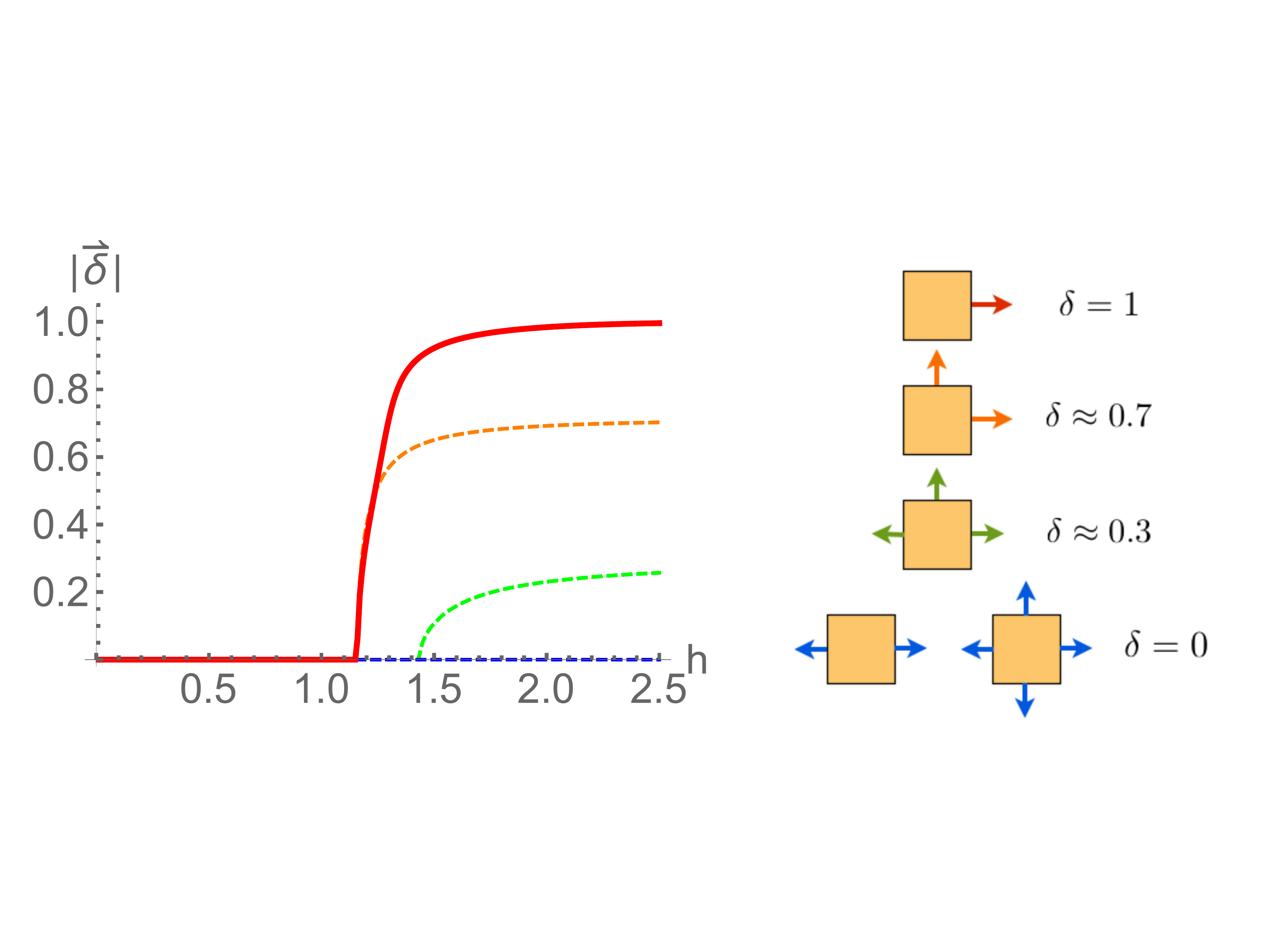}
\caption{Left: $| \vec{\delta} |$ in translation-invariant steady states as a function of 
the Hill exponent $h$ in the 
low diffusion regime ($D=10^{-7}$ $\mu\text{m}^2 / \text{s}$). Stable steady states are
shown in red, the others are linearly unstable. Right: for each type of
steady state, we show the corresponding PIN configurations along with
the norm of $\delta$ in the limit of large $h$.
}
\label{fig:bifurcation2D}
\end{figure*}

\change{To reveal the importance of the parameter $h$, we show the behavior of the polarization in 
Fig.~8. Just as in one dimension}{Equivalently}, for low values of $h$ there is
only one steady state and it is unpolarized \add{(cf. Fig.~8)}. 
Increasing $h$, there is spontaneous symmetry breaking
at a first threshold where the unpolarized state becomes unstable and a new 
polarized steady state appears. Cells in that polarized state have a 
\change{greatest}{large} number of PIN transporters
on one face and no polarization in the perpendicular direction. Because of this 
last property, the system effectively behaves
as a stack of rows which do not
exchange auxin, each row being like the polarized one-dimensional system.
Surprisingly, a second spontaneous symmetry breaking transition arises at 
\change{still}{a very slightly} larger value of $h$ \add{and even a third
still beyond that. The associated}
\remove{two other} translation-invariant steady states \change{appear}{behave}
as illustrated in Fig.~\ref{fig:bifurcation2D}. However these \change{new}{spurious} states
are always linearly unstable and so will not be considered further.

To get a global view of the \change{phase diagram}{behaviour} as a function of both $D$ and $h$,
we present via a heat map the complete phase diagram in Fig.~\ref{fig:doublephasevsDA}b
where the norm of the polarization vector is given only for the (unique)
stable (and translation-invariant) steady state.

\subsection*{\remove{Non translation-invariant steady states}}

\remove{Just as in the one-dimensional model, in the limit of large $h$,
each cell can polarize independently, so one has $2^M$ states if there
are $M$ cells. However as $h$ is lowered, far fewer steady states exist because
nearby cells tend to align their polarizations. To demonstrate this, we again follow
the procedure introduced in the one-dimensional model which allowed us to
follow the polarization of one or two defect cells in an otherwise
homogenous system. However, in the present two-dimensional case, 
there are two possibilities: the defect can
be polarized in the direction opposite to that of its neighbors as in 
one dimension, or it can be perpendicular to that direction. Either way,
the behavior is similar to what happens in the one-dimensional model, as illustrated in
Fig. 9 for the case of perpendicular
polarizations.}

\remove{One can speculate that the larger the 
domain of cells having similarly oriented polarizations, the greater the stability
of the associated steady state and the larger the size of its basin of attraction.
Thus for $D$ close to $D_c$, the dominant steady states (for instance as obtained
from relaxing from random initial conditions) should exhibit some local ordering of
their polarization patterns. Furthermore, it is plausible that the presence of noise
will enhance ordering as it does in one dimension, except that here
true long-range order may be possible.}

\subsection{Properties of the stochastic model}

\begin{figure}
\centering
\includegraphics[scale=0.4]{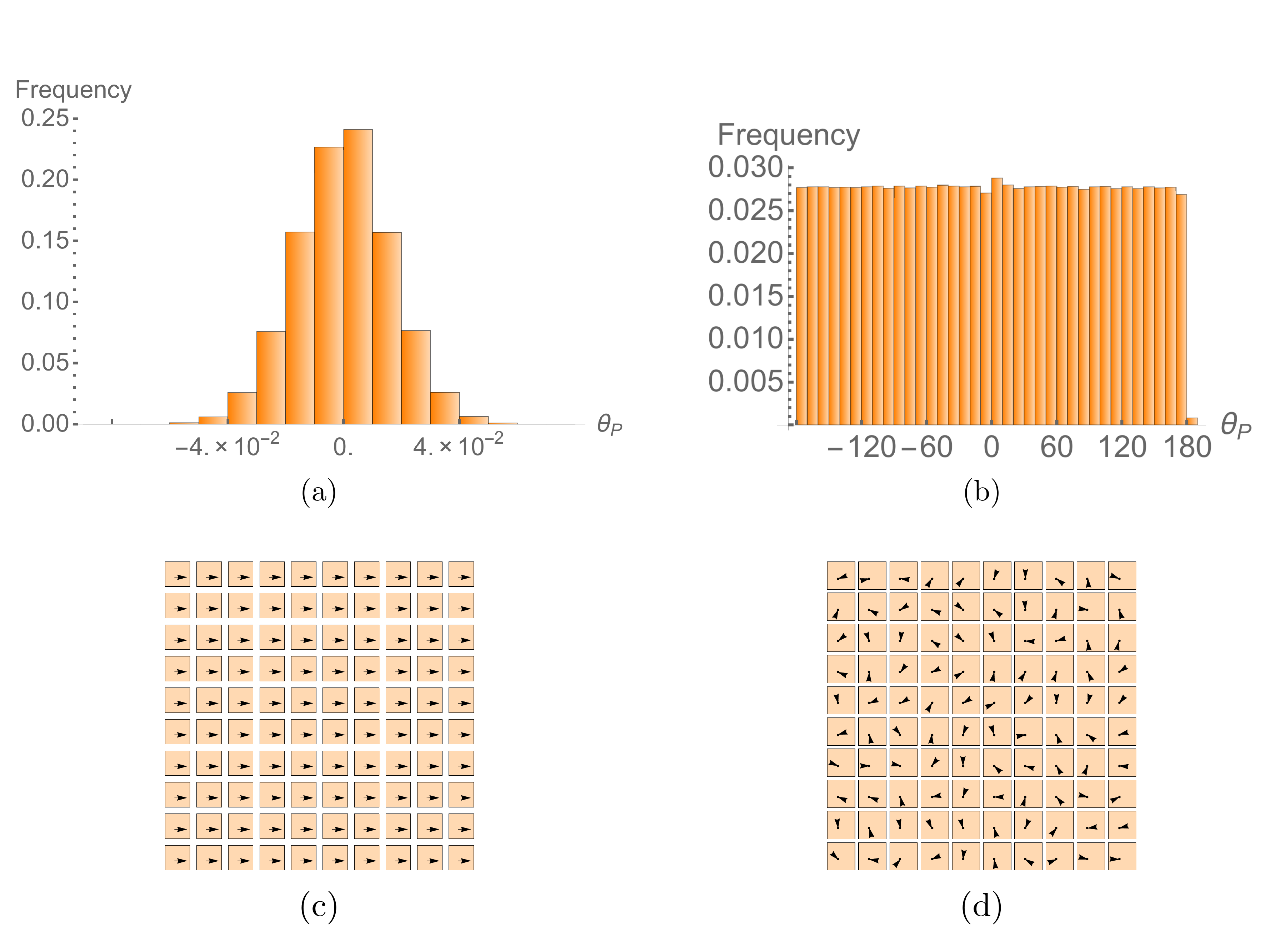}
\caption{Top: histograms of the lattice-wide polarization angle, $\theta_P$, in degrees, for
the regimes of 
low (left) and high (right) $D$, accumulated during a simulation of the model using
a 10 $\times$ 10 lattice. 
Bottom: typical equilibrium configurations in those same two regimes.
Left: $D=10^{-7}$ $\mu\text{m}^2 / \text{s}$. 
Right: $D=10^{-6}$ $\mu\text{m}^2 / \text{s}$.
}
\label{fig:ergodicityandconfigurations}
\end{figure}

The method of introducing molecular noise into the dynamical equations is oblivious
to the dimensionality of the model. Thus each dynamical equation can be rendered
stochastic for the two-dimensional model without any further thought by following the procedure
\change{previously described}{outlined above} for the one-dimensional case.
\remove{just consider each underlying 
flux and go from the average number of molecules transferred in the time step $dt$ to
a random number of molecules generated according to a Poisson law with the deterministic average. Ergodicity
of the dynamics follows and thus the uniqueness of the ``thermodynamic equilibrium state''.}
We \change{now}{can then use this to} study \change{that}{the thermodynamic equilibrium} state.
\remove{by simulation on a lattice with periodic boundary conditions.}
\remove{Initializing the PIN variables to be fully polarized, we thermalize the system and then
sample the thermodynamic equilibrium state. To test whether equilibrium is reached,
we also used unpolarized initial states and checked whether the initial condition
affects the averages obtained from the simulations. Just as in the one-dimensional 
case at low $D$, we found that 
the use of polarized initial conditions led to better thermalization than
the use of unpolarized initial conditions.}
Once equilibration was observed, we measured the average polarization vector
$\langle \vec{\delta} \rangle$, the average $\langle \cdot \rangle$ being taken
over the whole lattice at one specific time. We also define $\theta_P$
as the angle of that averaged vector, $\tan(\theta_P) = \delta_2/\delta_1$.
In the low $D$ regime, the cells stay highly
polarized and are oriented close to a common direction along one of
the axes of the lattice. This situation
illustrated in Fig.~\ref{fig:ergodicityandconfigurations} where we also show the
distribution of $\theta_P$ over the time of the simulation.
On the contrary, for ``high'' $D$, PINs tend to distribute quite evenly amongst
the faces of a cell and this leads to a relatively flat histogram
for the values of $\theta_P$ (Fig.~\ref{fig:ergodicityandconfigurations}).
However, \change{that}{this} histogram is a \change{bit}{slightly} misleading because the polarization vectors 
$\langle \vec{\delta} \rangle$ have a very
small magnitude and in effect each cell is \change{relatively}{essentially} depolarized. 

Just as in the one-dimensional case, one may ask whether there is
a true transition from a globally polarized state to an unpolarized
state when $D$ goes from low to high values. A naive way to do so would
be to average $\langle \vec{\delta} \rangle$ over the length of the simulation.
However, because the dynamics is ergodic, this average should vanish in the 
limit of a long run. The same difficulty arises in all systems that
undergo spontaneous symmetry breaking. It is necessary to first take the
norm of $\langle \vec{\delta} \rangle$, then average over time, and finally
check for trends with the size of the lattice. In Fig.~\ref{fig:phasevsDA}h
we show this time average, $\overline{| \langle \vec{\delta} \rangle|}$,
as a function of $D$ for lattices of different sizes. For comparison,
we also show the corresponding curve in the absence of noise. 

The behavior displayed is compatible
with a true ordering transition as might be expected from the analogy with 
the behavior of the Ising model. Such a behavior is also 
in agreement with the noise-induced ordering scenario~\cite{matsumoto1983} and related 
phenomena \cite{helbing2002}.

\section{Conclusions}
Although auxin transport in meristematic tissues (roots, shoots and cambium)
has been actively studied in the past decade while
associated molecular actors have been identified, the questions of
how intra-cellular PIN polarisation arises and how globally coherent
polarization patterns emerge have not been sufficiently addressed. Our 
work is based on modeling both auxin transport across cells and PIN recycling within 
individual cells. The dynamics we use for PIN recycling 
is modulated by an auxin flux-sensing system. Such recycling
allows PIN transporters to move within a cell from one face to
another. The PINs can accumulate on one face if there is a feedback
which allows \remove{for}\change{a polarization}{such a polarized state} to maintain itself.
Given this framework and estimates for a number of 
model parameter values, we mapped out a phase diagram giving the 
\remove{system's} behavior \add{of the system} in terms of specific parameters. The one-dimensional
model, describing a row of cells in a \change{meristem}{plant tissue}, allowed for large scale PIN 
polarization in the absence of any auxin gradient. Furthermore that
toy limit was analytically tractable and correctly described 
all the features arising in the two-dimensional
model. \change{Such a control of the model's properties}{The detailed analysis} revealed a
particularly essential ingredient: PIN polarization requires
a sufficient level of \emph{non-linearity} 
in the PIN recycling rates. In terms of our mathematical equations, this 
\change{cooperativity}{non-linearity} was parametrized by the Hill exponent $h$ appearing
in Eq.~\ref{eq:pinequation}, \add{which is associated with co-operativity in
the field of enzyme kinetics}. If \remove{there is
no cooperativity as in} Michaelis-Menten dynamics 
\add{is used} (corresponding to $h=1$ \add{and thus no 
co-operativity}), 
the system always goes to the unpolarized state\remove{preventing morphogenesis}.
\change{But}{On the other hand,} when $h$ rises above a threshold $h_c$ the homogeneous unpolarized
state becomes unstable and polarized PIN patterns spontaneously emerge.
We showed that the same qualitative behavior occurs when using non-linearities
based on stretched exponentials rather than Hill equations 
(\emph{cf.} Supplementary Material). That result shows that our model's predictions
are robust to changes in assumptions about the dynamical equations.

In addition, by studying the feedback between auxin concentrations
and PIN recycling, we showed that nearby cells tend to polarize 
in the same direction. 
Another particularly striking result found
was that the molecular noise in the PIN recycling dynamics seems to 
impose long range order on the PIN polarization patterns. This ``noise-induced
ordering'' could be the mechanism
driving the ordering found for instance in the cambium, ordering that can span 
tens of meters in the case of trees. 

Given that these conclusions follow from our hypothesis that 
PIN recycling is based on flux sensing, experimental investigations
should be performed to provide stringent comparisons with \add{the predictions of} our model\remove{'s predictions}.
The most direct test of our hypothesis would be to determine whether cells depolarize when
the auxin flux carried by PINs is suppressed. In \emph{Arabidopsis}, the polarization of PIN
can be observed thanks to fluorescent PIN transporters so what needs to be done
is to apply a perturbation affecting auxin flux. One simple way to \change{do so}{achieve this} is to 
inject auxin into an apoplast; the associated increase in auxin concentration will likely inhibit
PIN transport into that apoplast. If such an injection cannot be \change{done}{performed} without 
mechanically disrupting the cell membranes, a less invasive manipulation could be obtained
if \remove{it were possible to 
modify} the AUX1 transporters \add{can be modified} so that they may be locally photo-inhibited. Exposure
to a laser beam would then prevent the auxin from leaving a given apoplast, followed
by a rapid increase in auxin concentration just as in the simpler experiment previously proposed.
\remove{Note that }In both cases, our model predicts
that the PIN recycling dynamics would lead to depolarization of the
cell polarized \emph{towards} the apoplast while the neighboring cell, 
polarized away from the apoplast, would hardly be affected. 

\section{Acknowledgements}
We thank Fabrice Besnard and Silvio Franz for critical insights and Barbara Bravi and Bela Mulder for comments. 
This work was supported by the Marie Curie Training Network NETADIS (FP7, grant 290038).

\appendix{Supplementary Material to: \change{Modeling the emergence of polarity patterns in meristemic auxin transport}{Modeling the emergence of polarity patterns for the intercellular transport of auxin in plants}}

\section{Linearized auxin dynamics for given PIN configurations}

In the Main Text we determined auxin steady states assuming given 
\emph{translation-invariant} states for PIN. Given such PIN polarizations,
we balanced the synthesis and catabolism rates of auxin,
concluding that the steady-state concentration of auxin in cells was $A_c = \beta / \rho$.
Furthermore, as can be seen in \change{Figure 10}{Figure 7} of the Main Text \add{for the case of two dimensions},
the concentrations in the horizontal and vertical apoplasts have hardly
any dependence on PIN polarization \add{and in fact no such dependence
  at all in the one dimensional case}. These features suggest that auxin 
steady-state concentrations might be only \emph{weakly} dependent on the PIN polarizations,
whether these are translation-invariant or not, in which case the
linearization of the equations should suffice for an accurate
description of auxin dynamics. 

In this linearization approach, we take PIN polarizations to be arbitrary
but fixed in time, and we introduce the (small) deviations of auxin concentrations
compared to a reference state. That reference state can be arbitrary, but
for tractability, it will be taken to be translation invariant. Then the study of the
linearized equations will provide insights on (i) relaxational 
dynamics, (ii) the ``ferromagnetic'' coupling between nearest neighbour cells,
and (iii) how the auxin steady state
is affected by changes in PIN polarizations.

\subsection{Dynamical equations}

For pedagogical reasons, we provide the explicit equations only for the one dimensional
case; the equations for the two dimensional model are obtained using the same procedures.
In the (translation-invariant) reference state, we take auxin concentration
in cells (respectively apoplasts) to be $A_c^*$ (respectively $A_a^*$).  
Then we consider (small) deviations in auxin concentrations, \change{deviations
that}{which may vary from cell to cell or apoplast to apoplast}:
\begin{equation*}
A_c (x,t) = A^{*}_c + \delta A_c (x,t), \quad A_a (x, x+\Delta, t) = A^{*}_a + \delta A_a (x, x+\Delta,t)
\end{equation*}
where \add{$x$ labels cell position and} $\Delta=\Lambda+\lambda$ is the cell to cell distance \remove{as defined in the Main Text}. 
\add{Here we made the dependence on position explicit being $P \equiv x$.}
The dynamical equations (only for auxin, the PIN transporters being taken as fixed since
the recycling time is long) then become:
\begin{itemize}
\item for cells:
\begin{equation}
\begin{split}
\frac{d (A^*_c + \delta A_c(x,t))}{dt} & = \beta - \rho \Bigg[A^*_c+\delta A_c(x,t)\Bigg]\\
&+ \frac{D}{\Lambda \epsilon} \Bigg[2 A^*_a + \delta A_{a}(x-\Delta, x ,t) +   \delta A_{a}(x, x+\Delta,t) - 2 A^*_c - 2 \delta A_c(x,t)\Bigg]\\
&+ \frac{\alpha N^{AUX1}}{\Lambda^3} \Bigg[ \frac{A^*_a + \delta A_{a}(x-\Delta, x ,t)}{1+\frac{A^*_a + \delta A_{a}(x-\Delta, x ,t)}{A^*}+\frac{A^*_c+\delta A_c(x,t)}{A^{**}}}\Bigg]\\
&+ \frac{\alpha N^{AUX1}}{\Lambda^3} \Bigg[ \frac{A^*_a + \delta A_{a}(x, x+\Delta,t)}{1+\frac{A^*_a + \delta A_{a}(x, x+\Delta,t)}{A^*}+\frac{A^*_c+\delta A_c(x,t)}{A^{**}}}\Bigg]\\
&- \frac{\gamma}{\Lambda^3} \Bigg[ N^{PIN}_E(x) \frac{A^*_c+\delta A_c(x,t)}{1+\frac{A^*_{a} + \delta A_{a}(x-\Delta, x ,t)}{A^*}+\frac{A^*_c+\delta A_c(x,t)}{A^{**}}}\\
&+ N^{PIN}_W(x) \frac{A^*_c+\delta A_c(x,t)}{1+\frac{A^*_{a} + \delta A_{a}(x, x+\Delta, t)}{A^*}+\frac{A^*_c+\delta A_c(x,t)}{A^{**}}} \Bigg],
\end{split}
\label{eq:cellvariation}
\end{equation}
where $N^{PIN}_W(x)$ and $N^{PIN}_E(x)$ refer to the PIN transporters lying respectively on the left and right hand-side of the cell $C$ at position $x$.

\item for apoplasts:
\begin{equation}
\begin{split}
\frac{d (A^*_a + \delta A_a(x,x+\Delta,t))}{dt} & = \frac{D}{\lambda \epsilon} \Bigg[2 A^*_c + \delta A_c(x,t) + \delta A_c(x+\Delta,t) - 2 A^*_a - 2 \delta A_a (x, x+\Delta)\Bigg]\\
&- \frac{\alpha N^{AUX1}}{\lambda \Lambda^2} \Bigg[ \frac{A^*_a + \delta A_{a}(x, x+\Delta)}{1+\frac{A^*_a + \delta A_{a}(x, x+\Delta)}{A^*}+\frac{A^*_c+\delta A_c(x,t)}{A^{**}}}\\
&+\frac{A^*_a + \delta A_{a}(x, x+\Delta)}{1+\frac{A^*_a + \delta A_{a}(x, x+\Delta)}{A^*}+\frac{A^*_c+\delta A_c(x+\Delta,t)}{A^{**}}}\Bigg]\\
&+ \frac{\gamma}{\lambda \Lambda^2} \Bigg[ N^{PIN}_E (x) \frac{A^*_c+\delta A_c (x,t)}{1+\frac{A^*_a + \delta A_a(x, x+\Delta)}{A^*}+\frac{A^*_c+\delta A_c(x,t)}{A^{**}}}\\
&+ N^{PIN}_W (x+\Delta) \frac{A^*_c+\delta A_c(x+\Delta,t)}{1+\frac{A^*_{a} + \delta A_{a}(x, x+\Delta)}{A^*}+\frac{A^*_c+\delta A_c(x+\Delta,t)}{A^{**}}} \Bigg]
\end{split}
\label{eq:apovariation}
\end{equation} 
\end{itemize}
Linearizing for small variations $\delta A_c$ and $\delta A_a$, 
Eq.~\ref{eq:cellvariation} becomes:
\begin{equation}
\begin{split}
\frac{d \delta A_c(x,t)}{dt} &= \beta - \rho \Bigg[A^*_c + \delta A_c(x,t)\Bigg] + \frac{2D}{\Lambda \epsilon} \Bigg[A^*_a-A^*_c \Bigg] + \frac{D}{\Lambda \epsilon} \Bigg[\delta A_{a}(x-\Delta, x)+\delta A_{a}(x, x+\Delta) - 2 \delta A_c(x,t) \Bigg]\\
&+ \frac{\alpha N^{AUX1}}{\Lambda^3} \Bigg[ \frac{2 A^*_a}{1+\frac{A^*_a}{A^*}+\frac{A^*_c}{A^{**}}} - \frac{\frac{2 A^*_a}{A^{**}}}{(1+\frac{A^*_a}{A^*}+\frac{A^*_c}{A^{**}})^2} \delta A_c(x,t) \Bigg]\\
&+ \frac{\alpha N^{AUX1}}{\Lambda^3} \frac{1+\frac{A^*_c}{A^{**}}}{(1+\frac{A^*_a}{A^*}+\frac{A^*_c}{A^{**}})^2} \Bigg[\delta A_{a}(x, x+\Delta) + \delta A_{a}(x-\Delta, x) \Bigg]\\
&-\frac{\gamma \sigma}{\Lambda^3} \Bigg[ \frac{A^*_c}{1+\frac{A^*_a}{A^*}+\frac{A^*_c}{A^{**}}} + \frac{1+\frac{A^*_a}{A^*}}{(1+\frac{A^*_a}{A^*}+\frac{A^*_c}{A^{**}})^2} \delta A_c(x,t)\Bigg]\\
& + \frac{\gamma}{\Lambda^3} \frac{\frac{A^*_c}{A^*}}{(1+\frac{A^*_a}{A^{*}}+\frac{A^*_c}{A^{**}})^2} \Bigg[N^{PIN}_E (x) \delta A_{a}(x, x+\Delta) +N^{PIN}_W (x) \delta A_{a}(x-\Delta, x)\Bigg]
\end{split}
\end{equation}
where $\sigma$ is the total number of PINs in a cell ($\sigma$ is the same for all cells).
Two kinds of terms can be identified in the previous equation: \remove{there are} inhomogeneous terms and homogeneous terms (linear in $\delta A_a$ or $\delta A_c$). Gathering together all the terms belonging to each class, we obtain:
\begin{equation}
\frac{d \delta A_c(x,t)}{dt}=f_C+ g \delta A_c(x,t) + \Bigg(\Phi + b_{\pi(C^E)}\Bigg) \delta A_{a}(x, x+\Delta)+\Bigg(\Phi + b_{\pi(C^W)}\Bigg) \delta A_{a}(x-\Delta,x)
\label{eq:cell}
\end{equation}
where:
\begin{itemize}
\item $f_C=\beta-\rho A^*_c + \frac{2 D}{\Lambda \epsilon} (A^*_a - A^*_c) + \frac{2 \alpha N^{AUX1}}{\Lambda^3} \frac{A^*_a}{1+\frac{A^*_a}{A^*}+\frac{A^*_c}{A^{**}}}-\frac{\gamma \sigma}{\Lambda^3} \frac{A^*_c}{1+\frac{A^*_a}{A^*}+\frac{A^*_c}{A^{**}}}$;
\item $g=-\rho-\frac{2 D}{\Lambda \epsilon} - \frac{2 \alpha N^{AUX1}}{\Lambda^3} \frac{\frac{A^*_a}{A^{**}}}{(1+\frac{A^*_a}{A^*}+\frac{A^*_c}{A^{**}})^2}-\frac{\gamma \sigma}{\Lambda^3} \frac{1+\frac{A^*_a}{A^{*}}}{(1+\frac{A^*_a}{A^*}+\frac{A^*_c}{A^{**}})^2}$;
\item $\Phi=\frac{D}{\Lambda \epsilon} + \frac{\alpha N^{AUX1}}{\Lambda^3} \frac{1+\frac{A^*_c}{A^{**}}}{(1+\frac{A^*_a}{A^*}+\frac{A^*_c}{A^{**}})^2}$;
\item $b_{\pi({C^s})}=\frac{\gamma N^{PIN}_s}{\Lambda^3} \frac{\frac{A^*_c}{A^*}}{(1+\frac{A^*_a}{A^*}+\frac{A^*_c}{A^{**}})^2}$, where $\pi(C^s)$ shows the dependence of $b$ on the polarisation $\pi$ of the face $s$ of the cell $C$, \add{$s=E$ or $W$ for right and left}.
\end{itemize}
Note that the inhomogeneous term, $f_C$, vanishes if the reference state is in
fact a steady state.

Proceeding in the same way for apoplasts, Eq.~\ref{eq:apovariation} becomes:
\begin{equation}
\begin{split}
\lambda \frac{d \delta A_a(x,x+\Delta,t)}{dt} &=f_a + \Bigg[p+ p_{\pi({C^E})} + p_{\pi({C^W})}\Bigg] \delta A_a(x, x+\Delta) + \Bigg[q_{\pi({C^E})} \delta A_c (x+\Delta,t) + d_{\pi({C^E})}\Bigg]\\
&+ \Bigg[q_{\pi({C^W})} \delta A_c (x,t) + d_{\pi({C^W})}\Bigg],
\end{split}
\label{eq:aposteady}
\end{equation}
where:
\begin{itemize}
\item $f_a=2 \frac{D}{\epsilon} (A^*_c-A^*_a) -\frac{2 \alpha N^{AUX1}}{\Lambda^2}\frac{A^*_a}{1+\frac{A^*_a}{A^*}+\frac{A^*_c}{A^{**}}}$;
\item $p=-2 \frac{D}{\epsilon} - 2 \frac{\alpha N^{AUX1}}{\Lambda^2} \frac{1+\frac{A^*_c}{A^{**}}}{(1+\frac{A^*_a}{A^*}+\frac{A^*_c}{A^{**}})}$;
\item $p_{\pi({C^s})}=-\frac{\gamma}{\Lambda^2} N^{PIN}_s \frac{\frac{A^*_c}{A^*}}{(1+\frac{A^*_a}{A^*}+\frac{A^*_c}{A^{**}})^2}$;
\item $q_{\pi({C^s})}=\frac{D}{\epsilon}+\frac{\alpha N^{AUX1}}{\Lambda^2} \frac{\frac{A^*_a}{A^{**}}}{(1+\frac{A^*_a}{A^*}+\frac{A^*_c}{A^{**}})^2}+\frac{\gamma}{\Lambda^2} N^{PIN}_s \frac{1+\frac{A^*_a}{A^*}}{(1+\frac{A^*_a}{A^*}+\frac{A^*_c}{A^{**}})^2}$;
\item $d_{\pi({C^s})}=\frac{\gamma}{\Lambda^2} N^{PIN}_s \frac{A^*_c}{1+\frac{A^*_a}{A^*}+
\frac{A^*_c}{A^{**}}}$.
\end{itemize}
To simplify these systems of equations, we note that $\lambda \ll \Lambda$ so that the time scale
for change of auxin concentrations in apoplasts is far shorter than that in cells. The simple reason
is that apoplasts have much smaller volumes so concentrations there change 
faster when subject to a given flux. We shall
thus consider that $\delta A_a$ is a fast variable in quasi equilibrium with the
slower variations of $\delta A_c$. In the limit $\lambda / \Lambda \rightarrow 0$, 
the left-hand side of Eq.~\ref{eq:aposteady} vanishes. Thus the right-hand side 
of that equation also vanishes, leading to an equation that
allows us to express $\delta A_a(x,x+\Delta,t)$ in terms of 
$\delta A_c (x,t)$ and $\delta A_c (x+\Delta,t)$.
Substituting this expression into Eq.~\ref{eq:cell} leads to the following equation
for intracellular auxin concentrations: 
\begin{equation}
\begin{split}
\frac{d \delta A_c(x,t)}{dt}&=f_C+g \delta A_c-\sum_{C^s} \Bigg[\Phi +b_{\pi({C^s})}\Bigg] \Bigg[ \frac{f_a + d_{\pi({C^s})} + d_{\pi({C'^{s'}})}}{p+p_{\pi({C^s})}+p_{\pi({C'^{s'}})}}\\
&+ \frac{q_{\pi(C^s)} \delta A_c}{p+p_{\pi({C^s})}+p_{\pi({C'^{s'}})}} + \frac{q_{\pi({C'^{s'}})} \delta A_{C'^{s'}}}{p+p_{\pi({C^s})}+p_{\pi({C'^{s'}})}}\Bigg]
\end{split}
\label{eq:final}
\end{equation}
where we use the notation $C^{s}$ to refer to the face $s$ of the cell $C$ at $x$, while
$C'^{s'}$ refers to the associated face $s'$ of cell $C'$, $C^{s}$ and $C'^{s'}$ delimiting the
apoplast separating $C$ and $C'$.

\subsection{Cell-cell couplings are ferromagnetic}

It is possible to draw a correspondence between Eq.~\ref{eq:final} and the dynamics
in standard many\add{-}body problems. Consider a lattice of continuous
variables $\sigma_i$, $i=1, 2, ... $, whose dynamics is \change{ruled}{governed} by the 
Hamiltonian $\mathcal{H}(\{ \sigma_i \})$:
\begin{equation}
\mathcal{H}(\{ \sigma_i \})=-\sum_i \sum_j J_{ij} \sigma_i \sigma_j - \sum_{i} h_i \sigma_i,
\end{equation}
where $J_{ij}$ is the coupling between the $i$-th and the $j$-th variable
while $h_i$ can be thought of a position-dependent external field.
In most physical systems, the matrix $J$
couples only nearest neighbours and can have diagonal elements, just as in Eq.~\ref{eq:final}.
The dynamics of the $\sigma$ variables are given by
\begin{equation}
\frac{d \sigma_i}{dt}=-\frac{\delta \mathcal H}{\delta \sigma_i} = \sum_j J_{ij} \sigma_j + h_i
\end{equation}
Let us now identify the $\sigma$s with the $\delta A_c$s. The  
set of equations obtained from Eq.~\ref{eq:final} by running over
all $N_{cell}$ cells can be written as an equation for the vector $\vec{\delta A_c}$:
\begin{equation}
\frac{\vec{\delta A_c}}{dt}=J \vec{\delta A_c} + \vec{h}
\label{eq:final-vector}
\end{equation}
where $J$ is an $N_{cell} \times N_{cell}$ coupling matrix while 
$\vec{\delta A_c}$ and $\vec{h}$ are $N_{cell}$ component vectors.

Using Eq.~\ref{eq:final}, the effective couplings between two nearest neighbour cells $C$ and 
$C'$ arises through the faces $s$ and $s'$ delimiting the apoplast 
between them, leading to:
\begin{equation}
J_{C-C'}=\Bigg[\Phi + b_{\pi(C^s)}\Bigg] \frac{q_{\pi(C'^{s'})}}{p+p_{\pi(C^s)}+p_{\pi(C'^{s'})}},
\end{equation}
Similarly, we find for the self-coupling:
\begin{equation}
J_{C-C}=g-\sum_{C^s}\Bigg[\Phi +b_{\pi({C^s})}\Bigg] \frac{q_{\pi(C^{s})}}{p+p_{\pi(C^s)}+p_{\pi(C'^{s'})}}
\end{equation}
\change{Lastly}{Finally}, the components of the field vector $\vec{h}$ are given by:
\begin{equation}
h_C=f_C-\sum_{C^s}\Bigg[\Phi + b_{\pi(C^s)} \Bigg] 
\frac{f_a+d_{\pi(C^{s})}+d_{\pi(C'^{s'})}}{p+p_{\pi(C^s)}+p_{\pi(C'^{s'})}}.
\end{equation}

To understand what the coupling matrix $J$ tells us
about the system, we set
$A_c^*$ and $A_a^*$ to their values in the steady state when PIN 
polarization is maximum and positive for all
cells. Given the linearization about
these values, the derivations \remove{of the previous section} show that
the matrix $J$ depends on the values of
PIN polarizations at each site, as it must. 

\begin{figure}[ht]
\centering
\includegraphics[scale=0.5]{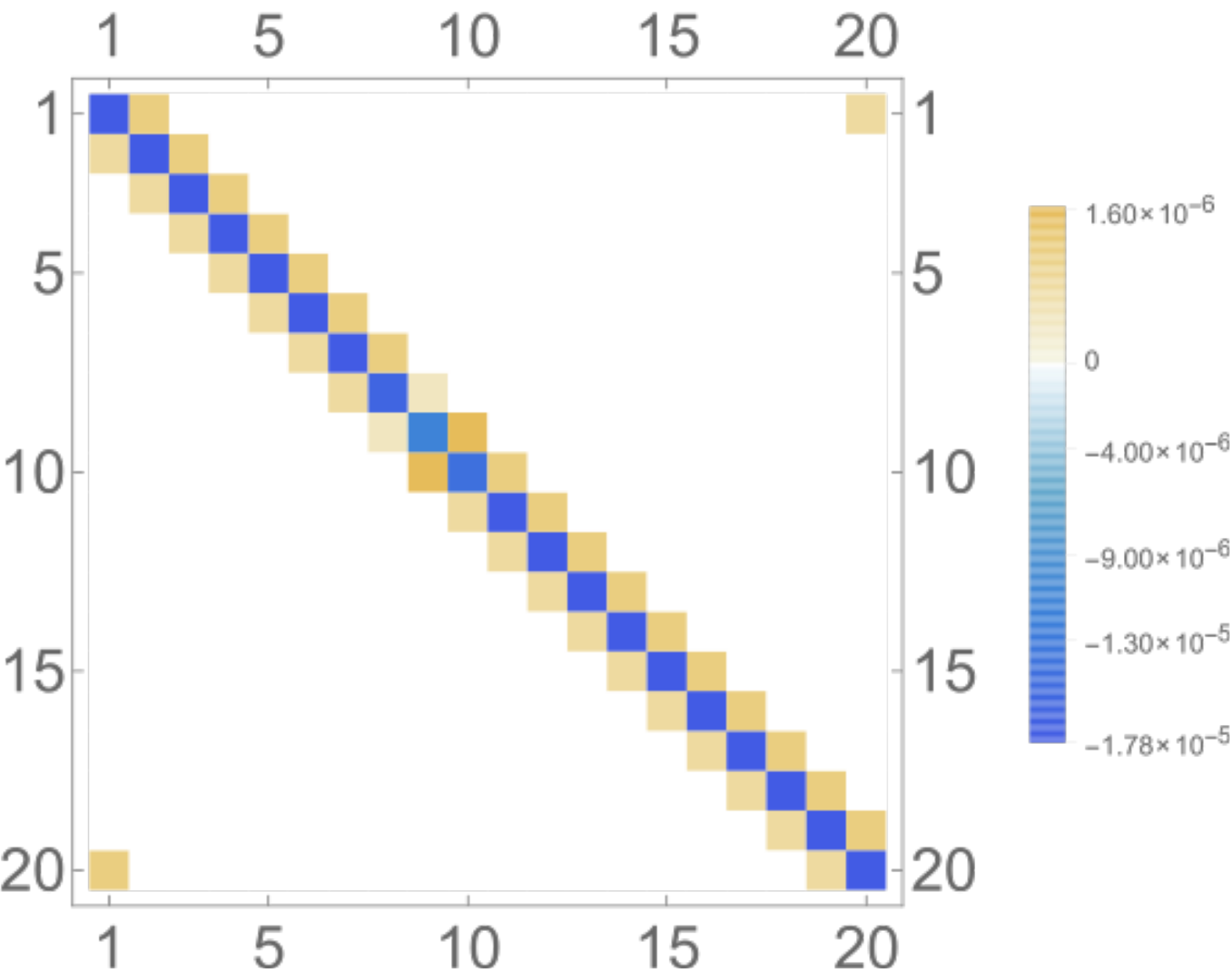}
\caption{Heat map of the matrix $J$ obtained for 20 cells in a row that are 
all maximally polarized in the plus direction but where the cell number 9 is
oppositely polarized. $D=10^{-7}$ $\mu\text{m}^2 / \text{s}$, all other parameters 
are set as in Table I of the Main Text.}
\label{fig:matrix1D}
\end{figure}

We have computed $J$ numerically for a lattice of 20 cells and
periodic boundary conditions where all cells had PIN polarization
equal to +1 except the cell at position 9 which had PIN polarization equal
to -1. Cell number 9 thus corresponds to a defect. Far away from the defect, one has the same situation as if the defect were absent,
and the matrix $J$ is translation invariant. One also has $J_{C-C} < 0$
which is expected given the stability of the steady state: indeed, if one
sets $\delta A_c(C,t)$ to a $C$-independent positive value, the 
$\delta A_c(C,t)$ should relax to zero with time. Furthermore, we find
$J_{C-C'} > 0$ for $C$ and $C'$ nearest neighbours,  
corresponding to a ferromagnetic interaction. Both of these
results are clearly visible in Fig.~\ref{fig:matrix1D} away from the defect. If 
auxin dynamics were just of the passive diffusion type, $J_{C-C'}$ would clearly be
positive (\emph{cf.} the linear diffusion equation), but we see that
neither the active transport processes nor the need to go from cell to cell via apoplasts
affect this conclusion. Interestingly,
the sign properties of the $J$ matrix elements
hold over the whole lattice and in particular around the defect
as can be seen in 
Fig.~\ref{fig:matrix1D}. Note that both the diagonal and off-diagonal elements 
of $J$ are slightly affected by the defect but not enough for the sign of
any element to change. 

Not surprisingly, the overall approach can be generalized to the two-dimensional case. The main difference \change{with}{to} the one-dimensional case lies in the distinction between the two types of apoplasts, \emph{i.e.}, the up and right hand-side apoplasts, and the fact that one has to deal
with the four PIN transporter numbers for each cell (one for each side). Following 
the same steps, one obtains the expression for the $N_{cell}\times N_{cell}$ matrix $J$
which couples nearest neighbour cells. 
The diagonal elements of $J$ are \remove{again} negative while the off diagonal ones are positive, 
again indicating effective ferromagnetic couplings between cells.

\begin{figure}[ht]
\centering
\includegraphics[scale=0.5]{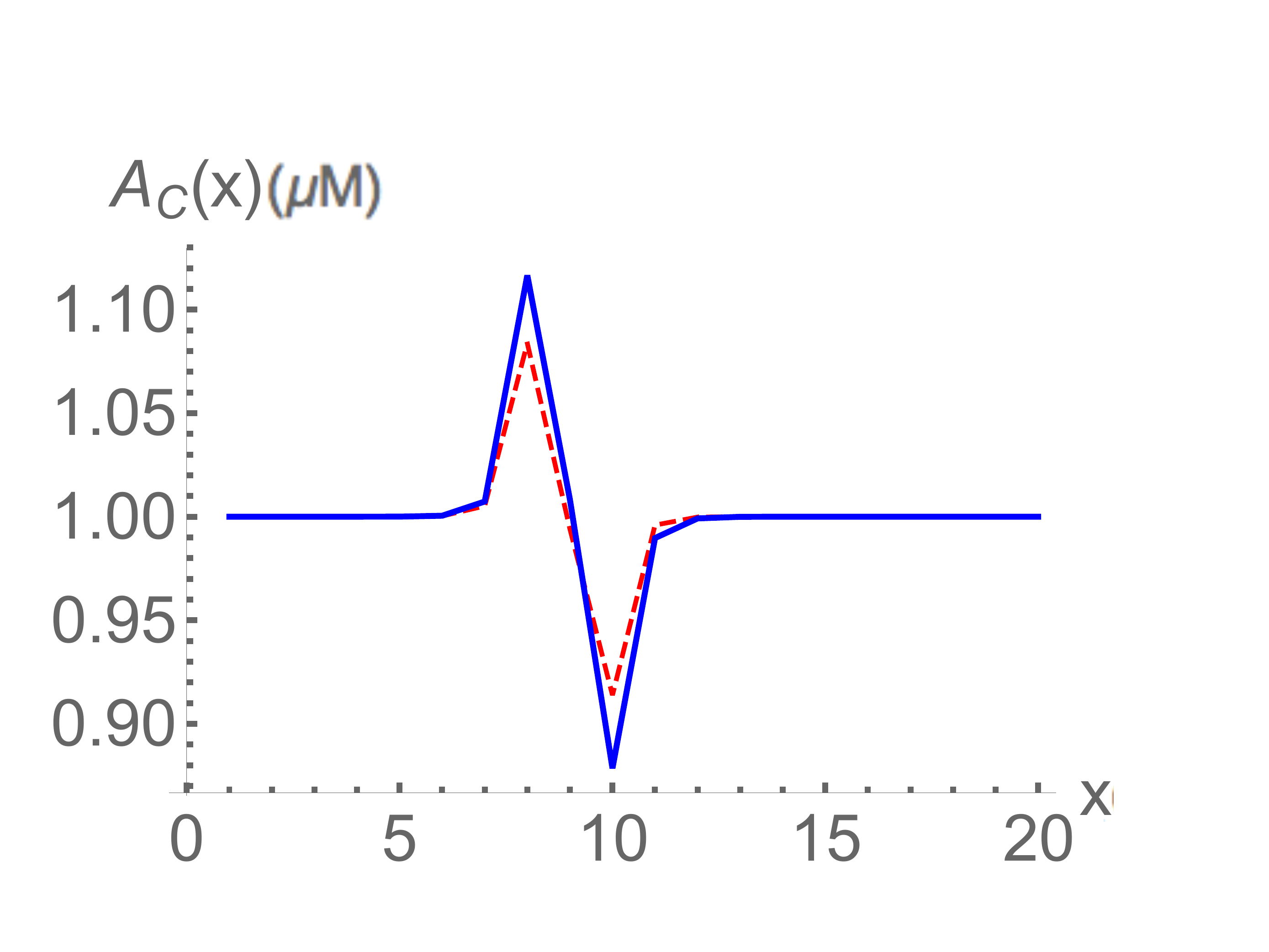}
\caption{Auxin concentration in cells as a function of the position along the chain. Blue thick line: prediction of the linearized dynamics. Red dashed line: results from simulation. There is one defect cell at position 9.  $D=10^{-7}$ $\mu\text{m}^2 / \text{s}$. All other parameters are set as in Table I in the Main Text. The presence of the defect creates an increase of auxin concentration on its left and a depletion on its right. The linearized approach is qualitatively good but is less precise near the defect.}
\label{fig:auxinvariation}
\end{figure}

\change{Lastly}{Finally}, the linearized dynamics (Eqs.~\ref{eq:final} or \ref{eq:final-vector}) can also be used 
to probe the steady state auxin concentrations for arbitrary PIN configurations.
For illustration, consider again the one-dimensional system of 20 cells
that are maximally positively polarized except for cell 9 which has opposite polarization. 
The \change{steady}{\emph{steady}}\add{}\change{state}{\emph{state}} in the linearized approximation corresponds to setting
the left-hand side of Eq.~\ref{eq:final-vector} to zero. The resulting equation
gives an estimate of the auxin concentration in each cell; \add{the} result \remove{that} is displayed in
Fig.~\ref{fig:auxinvariation}. By comparing to the exact concentrations
(using the full non-linear equations), we see that the linear approximation
provides a qualitatively satisfactory description of the changes \add{in auxin concentrations} induced
by the defect.

\section{\add{Energy-like function for the translation-invariant PIN dynamics in the quasi-equilibrium limit for auxin}}

\add{As discussed in the Main Text, in Sec. IIIC, assuming translation-invariance, PIN dynamics follow from a potential energy function. Auxin concentrations are time independent. Considering this framework and the constraint on PIN number conservation, the time derivative of the polarization $\delta$ is proportional to the gradient of the energy function $\mathcal{F}(\delta)$ (Eq. 9 in the Main Text), i.e., $\frac{d\delta}{dt}=-\frac{1}{\tau_{1D}} \frac{d \mathcal{F}}{d\delta}$. To obtain the closed form for $\mathcal{F}$ it is enough to integrate with respect to $\delta$ the right-hand side of the differential equation. For $h=2$ the integral can be performed in closed form, giving:}

\begin{equation}
\mathcal{F}(\delta)=\frac{\sigma^2}{2 c^2} \mbox{log}\Big[1+\Big(\frac{c(1+\delta)}{2}\Big)^2\Big]+\frac{\sigma^2}{2 c^2} \mbox{log}\Big[1+\Big(\frac{c(1-\delta)}{2}\Big)^2\Big],
\end{equation}

\add{where $c=\frac{\gamma \sigma}{\phi^*}\frac{A_c}{1+\frac{A_a}{A^*}+\frac{A_c}{A^{**}}}$. One can check that $c$ grows with the inverse of the diffusion constant; indeed $A_c$ does not depend on $D$ and $A_a$ grows with increasing $D$ (Fig. 2 in the Main Text).}

\add{For any value of $h \neq 1$ $\mathcal{F}(\delta)$ is given in terms of hypergeometric functions, $_2 F_1$}\cite{integral,integral2}\add{:}

\begin{equation}
\mathcal{F}(\delta) =\frac{\sigma^2 (1+\delta)^2}{4 \tau_1} \cdot _2F_1\bigg[1, \frac{2}{h}, 1+\frac{2}{h}, -\bigg(\frac{c(1+\delta)}{2}\bigg)^h\bigg] + 
\frac{\sigma^2 (1-\delta)^2}{4 \tau_1} \cdot _2F_1\bigg[1, \frac{2}{h}, 1+\frac{2}{h}, -\bigg(\frac{c(1-\delta)}{2}\bigg)^h\bigg].
\end{equation}

\add{A series representation is $_2F_1[a,b,d,z]=\sum_{k=0}^{\infty} \frac{(a)_k (b)_k}{(d)_k k!} z^k$, $(a)_k$ being the Pochhammer symbol.}

\section{Critical line in the phase diagram}

In the Main Text we exploited the effective potential found in the one-dimensional model to
obtain the critical diffusion constant $D_c$. In direct analogy with that
calculation, one can obtain the equation for $h_c$, the critical value of the Hill exponent $h$.
The starting condition is:
\begin{equation}
\frac{\partial^2 \mathcal{F}}{\partial {N^{PIN} _E}^2} \bigg\vert_{N^{PIN}_E=\frac{\sigma}{2}}=0
\end{equation}
Setting $c=\frac{\gamma \sigma}{\phi^*} \frac{A_c}{1+\frac{A_a}{A^*}+\frac{A_c}{A^{**}}}$
which depends on $D$ through \change{the}{its} dependence on $A_a$, 
we obtain:
\begin{equation}
(h-1) \Bigg(\frac{c}{2}\Bigg)^h=1
\end{equation}
Applying the logarithm to both the sides, one gets:
\begin{equation}
h=-\frac{\log(h-1)}{\log(\frac{c}{2})}
\label{eq:critical_h}
\end{equation}
This equation provides the critical line in the $(D, h)$ plane 
(see Main Text, Fig. 6, green dashed line). 

\section{\add{Non translation-invariant steady states in the 1D Model}}

\add{If parameter values fall in the range where Fig.~6a  in the Main Text shows high spontaneous polarization,
random initial configurations will relax to steady states having quite random 
PIN polarizations. In effect, the coupling between neighboring cells is low in that regime
and one can expect to have $2^M$ steady states if there are $M$ cells.
The typical steady state is then disordered, with no coherent transport of auxin.
As $D$ increases or $h$ decreases, the number of these steady states decreases:
neighboring cells more often than not have the same sign of PIN polarization. 
To get some insight into this 
phenomenon, which seems analogous to ferromagnetism, 
let us consider what happens when a localized \textit{defect} is 
introduced into an otherwise uniformly
polarized system. Beginning with the uniformly polarized steady state
at low $D$, we reverse the polarization
of \emph{one} cell to form a defect and then we let the system relax to produce
an initial modified steady state. Thereafter we follow this steady state 
as we increase $D$. The resulting polarizations of the defect cell are
shown in Fig.~4c  in the Main Text. We also display what 
happens in the absence of the defect (red curve in Fig.~4c  in the Main Text). We see that
the defect cannot sustain itself arbitrarily close to the threshold $D_c$: when polarizations
are too weak, the defect cell will align 
its polarization with its
neighbors and the defect will disappear. The curve of defect polarization as a function of 
$D_c$ thus has a discontinuity at some value strictly lower than $D_c$
as can be seen in Fig.~4c  in the Main Text. The interpretation should be clear: of the
$2^M$ putative steady states with random polarizations, some in fact do not exist, and one can
expect the number of steady states to decrease steadily as one
approaches $D_c$. }

\add{To provide further evidence in favor of such a scenario, consider what happens when there 
are two defects next to one another as in Fig.~4e in the Main Text. We see that just
as for the single defect, two adjacent defects disappear strictly before $D_c$ 
\emph{but} they do so \emph{after} the single defect does. Another interesting feature
is that the two defects have slightly different polarizations. This asymmetry
is unexpected if one considers the analogy with the Ising model but in fact
it is unavoidable here: indeed, the apoplast on the left of the left defect is enriched in auxin while
the apoplast on the right of the right defect is depleted in auxin. We have checked
that this difference vanishes if the two single-cell defects are placed far away from one
another and also that in such a limit the polarization curves converge to those
of single defects.}

\section{\add{Non translation-invariant steady states in the 2D Model}}

\add{Just as in the one-dimensional model, in the limit of large $h$
each cell can polarize independently, so one has $2^M$ states if there
are $M$ cells. However as $h$ is lowered, far fewer steady states exist and
nearby cells tend to align their polarizations. To demonstrate this, we again follow
the procedure introduced in the one-dimensional model which allowed us to
follow the polarization of one or two defect cells in an otherwise
homogenous system. However, in the present two-dimensional case, 
there are two possibilities: the defect can
be polarized in the direction opposite to that of its neighbors as in 
one dimension, or it can be perpendicular to that direction. Either way,
the behavior is similar to what happens in the one-dimensional model, as illustrated in
Fig.~4d-e in the Main Text for the case of perpendicular
polarizations.

One can speculate that the larger the 
domain of cells having similarly oriented polarizations, the greater the stability
of the associated steady state and the larger the size of its basin of attraction.
Thus for $D$ close to $D_c$, the dominant steady states (for instance as obtained
from relaxing from random initial conditions) should exhibit some local ordering of
their polarization patterns. Furthermore, it is plausible that the presence of noise
will enhance ordering as it does in one dimension, except that
true long-range order may be possible in two dimensions.}

\section{\add{Going from the deterministic to the stochastic model}}

\add{To include molecular noise in 
reaction-diffusion systems, it is common practice to use a Lattice Boltzmann model framework [1] and thermodynamical considerations to render the reaction dynamics
stochastic. However, in the present case, the Michaelis-Menten form of the rates of change of 
molecular concentrations as well as the PIN recycling dynamics
do not correspond to mass action reactions so a different framework is necessary. We thus take a Gillespie-like 
approach [2] where each process type (production or degradation of auxin inside a cell, transfer
of auxin from one side to the other of a
membrane, or PIN recycling from one cell face to another) is rendered stochastic.
For instance, diffusion of auxin across a face of a cell is associated with two underlying unidirectional
fluxes, one in each direction. If $dt$ is a short time interval, the noiseless value of one of these fluxes
specifies that an average number of molecules ${\cal{M}} = N r dt$ will contribute to the 
underlying flux where $N$ is the number of molecules potentially concerned by the process and $r$ is the probability per unit time that a molecule will be transferred. Because of the molecular nature of the process,
the true number of molecules transferred will be a 
Poisson variable of mean $N r dt$. Applying this rule to all the 
underlying fluxes contributing to the auxin differential equations and to Eq. 7, the stochastic dynamics are naturally 
implemented without the introduction of any new parameter. (Note for instance
that temperature dependencies feed-in only via the deterministic parameters such as $D$.)
However, there are millions of auxin molecules in a cell and so the associated noise is
negligible. In contrast, the number of PIN transporters in a cell is modest and so
the noise in PIN recycling can a priori be important. Our simulation thus implements
the noise in PIN recycling but not in the other processes. The discretization in time using the time step $dt$ is exact only if all numbers of molecules are kept constant. The approach becomes exact, {\em i.e.} recovers the Gillespie method, when $dt$ goes to zero. In our work we check that $dt$ is small enough by comparing simulations at different values of $dt$. In practice we begin with a 
``cold start'', \emph{i.e.}, the starting configuration of the system has all cells similarly 
polarized; then we run to thermalize the system before performing any measurements.} 

\add{To check whether our simulations indeed produce equilibrium configurations, 
we compared the polarization values to those obtained when starting the simulations with
``hot starts'', \emph{i.e.}, the starting configuration of the system having cells randomly polarized.}
\add{For } $D \gt 5 \cdot 10^{-7}\text{m}^2/\text{s}$ \add{the two approaches agreed very well while for lower values of 
$D$ the runs with random initial conditions failed to thermalize well. As a result, only
for intermediate and large values of $D$ can one be confident in the simulation.}

\section{Polarisation Angle Susceptibility in the two-dimensional Model}

As a follow-up \change{of the last section of}{to the study of the two-dimensional model in the presence of noise in} the Main Text, it is of interest
to consider the variance of the polarisation angle as one approaches
the ordering transition at $D_c$. We thus examine the susceptibility of 
the angle $\theta_P$, \emph{i.e.}:
\begin{equation}
\chi_{\theta_P}=\overline{ \langle \theta^2_P \rangle - \langle \theta_P \rangle^2}
\end{equation}
where the average $\langle \cdot \rangle$ is taken over the lattice and the overline denotes the time average. 

\begin{figure}[ht]
\centering
\includegraphics[scale=0.5]{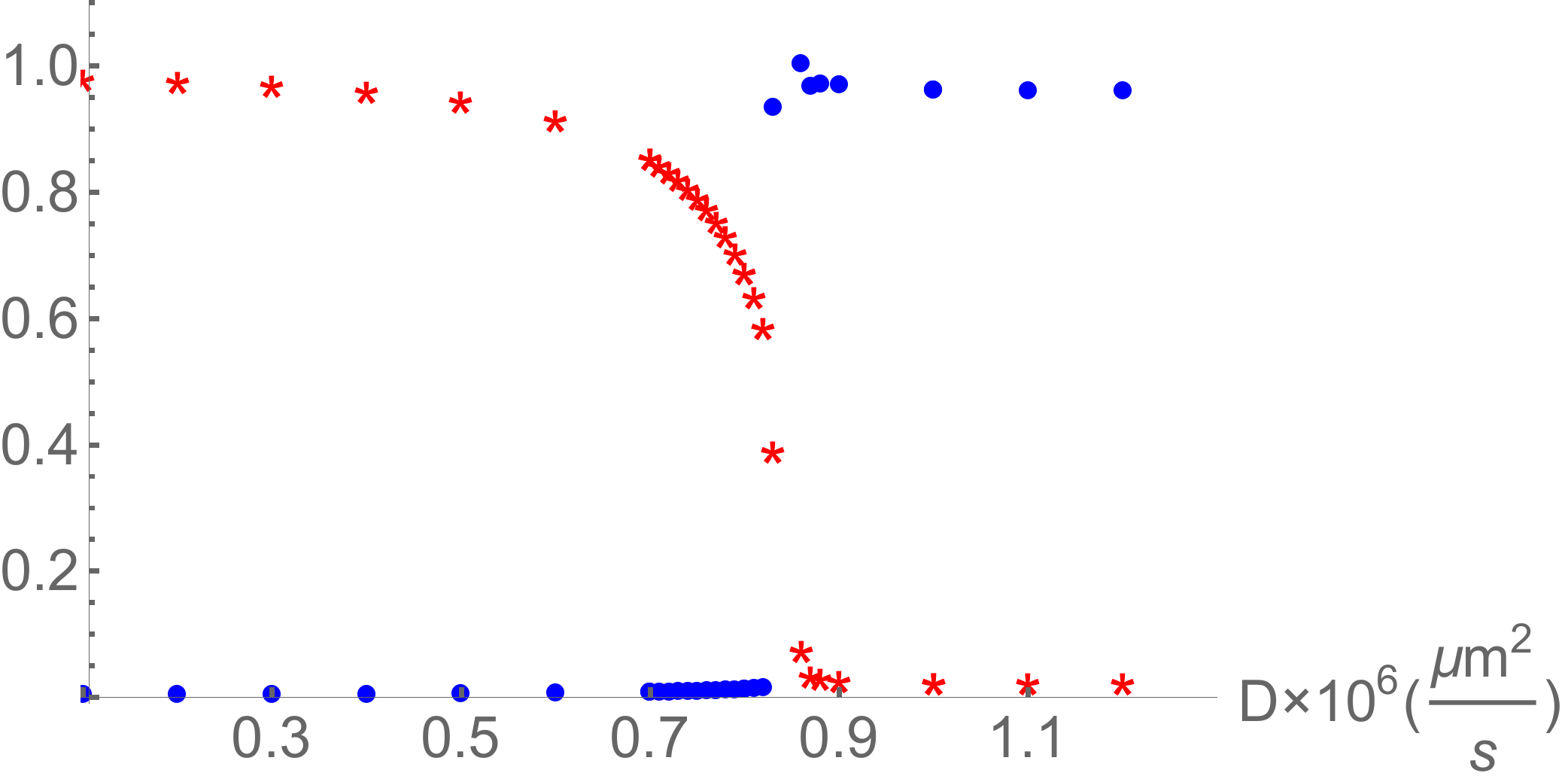}
\caption{Rescaled \change{squared}{square} root of the normalized angle susceptibility as defined in the text\add{,} as a function of diffusion constant (blue circles)\add{,} plotted along with the behaviour of the polarisation (red stars). The steep rise of the susceptibility takes place around the critical value of the diffusion constant as expected.}
\label{fig:susceptibility}
\end{figure}

In Fig.~\ref{fig:susceptibility} we show the square root of the susceptibility 
rescaled by its maximum \change{for different values}{as a function} of $D$ \add{for a 10$\times$10 lattice}. 
It stays close to zero (all the polarisation vectors pointing in a similar direction) 
in the low diffusion regime and then it rises dramatically as one
approaches $D_c$. Clearly once the disordered phase is reached, the 
variance is \change{maximum}{maximal} and no longer changes with $D$.
%

\section{\add{Replacing the Hill-equations by stretched exponentials in PIN recycling dynamics}}

\add{So far and throughout the Main Text, we have modeled PIN recycling as an enzymatic process, using a Hill function of the out-going flux, $\phi^{PIN}_{f}$:}

\begin{equation}
G(\phi^{PIN}_f)=\frac{1}{1+\Big(\frac{\phi^{PIN}_{f}}{\phi^* \Lambda^{-2}}\Big)^h}.
\end{equation}

\add{We found that the exponent $h$ plays a central role in the appearance of PIN polarization. However, one can ask whether other functional forms for PIN recycling give rise to an analogous result. To investigate this hypothesis, we performed the same bifurcation analysis as in the Main Text but using a stretched exponential instead of a Hill equation:}

\begin{equation}
G(\phi^{PIN}_f)=\mbox{e}^{-({\phi^{PIN}_f})^{\nu}},
\end{equation} 

\begin{figure}
\includegraphics[scale=0.5]{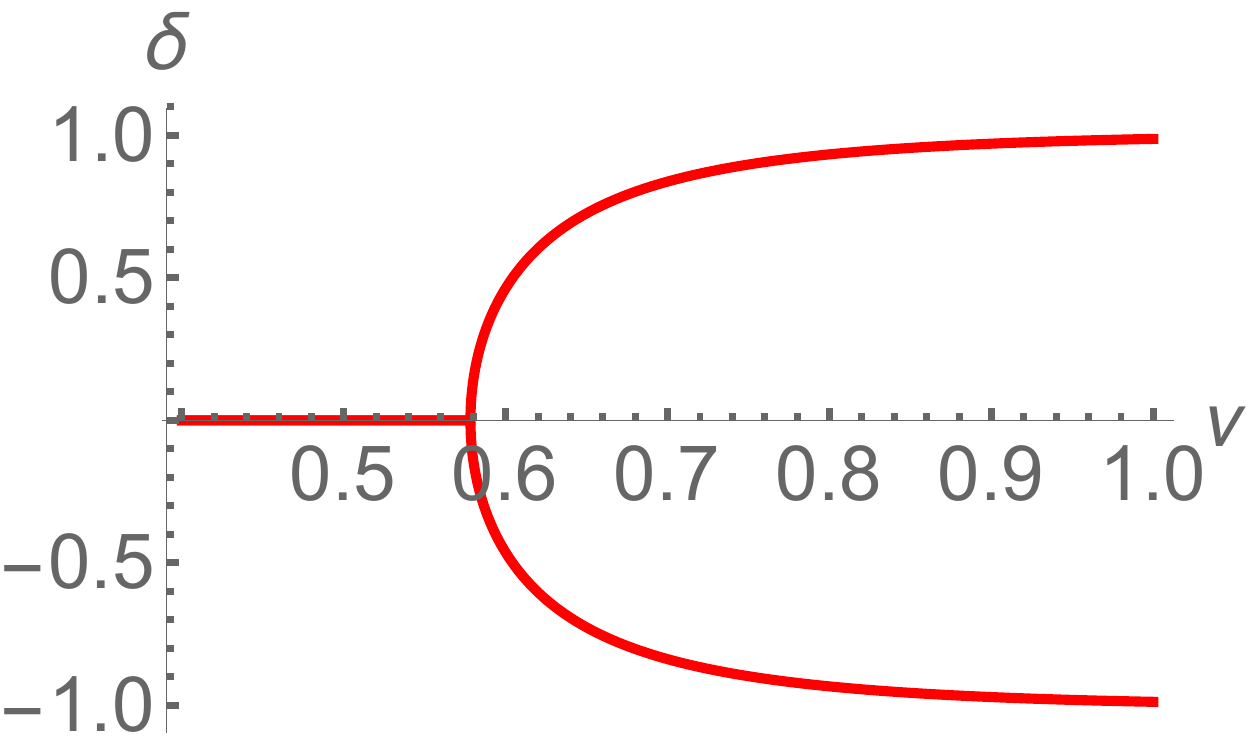}
\caption{Bifurcation diagram for translation-invariant states in the one-dimensional model. $\delta$ is the PIN polarization. The unpolarized state is stable for $\nu < \nu_C \simeq 0.58$ (red). Beyond this threshold, two symmetric polarized states appear. These are stable (in red) whereas the unpolarized state becomes unstable (not shown). Here $D = 10^{-7}$ $\frac{\mu m^2}{s}$ while other parameter values are given in Table I of the Main Text.}
\label{fig:bifurcationnew}
\end{figure}

\add{Such stretched exponentials are used to describe kinetics in a disordered physical system subject to collective effects} \cite{sornette}; \add{$\nu$ is an exponent playing a similar role as $h$ did in the Hill case. 

Using this non-linear function for $G(\phi^{PIN}_f)$, we have determined the phase diagram. Polarised states appear beyond $\nu \simeq 0.58$} (Fig. \ref{fig:bifurcationnew})\add{: above this critical value, the unpolarised state becomes unstable while the two polarised ones turn out to be stable as in the Hill framework. These behaviours suggest that the strength of the non-linearity in PIN recycling dynamics drives spontaneous polarisation.}

\end{document}